\documentclass[fleqn,usenatbib]{mnras}

\usepackage[T1]{fontenc}

\DeclareRobustCommand{\VAN}[3]{#2}
\let\VANthebibliography\thebibliography
\def\thebibliography{\DeclareRobustCommand{\VAN}[3]{##3}\VANthebibliography}

\usepackage{graphicx}	
\usepackage{amsmath}	
\usepackage{amssymb}	
\usepackage{lineno}
\usepackage{eso-pic}
\defcitealias{Wiseman2021}{W21}
\defcitealias{Kelsey2022}{K22}
\defcitealias{Brout2020}{BS21}
\defcitealias{Nicolas2020}{N21}
\usepackage{newtxtext,newtxmath}

\title[SN Ia host dust and age model]{ A galaxy-driven model of type Ia supernova luminosity variations}

\author[P. Wiseman et al.]{
\parbox{\textwidth}{
\Large
P.~Wiseman,$^{1}$\thanks{E-mail: P.S.Wiseman@soton.ac.uk}
M.~Vincenzi,$^{2,1}$
M.~Sullivan,$^{1}$
L.~Kelsey,$^{3,1}$
B.~Popovic,$^{2}$
B.~Rose,$^{2}$
D.~Brout,$^{4,5}$
T.~M.~Davis,$^{6}$
C.~Frohmaier,$^{3,1}$
L.~Galbany,$^{7,8}$
C.~Lidman,$^{9,10}$
A.~M\"oller,$^{11}$
D.~Scolnic,$^{2}$
M.~Smith,$^{12}$
M.~Aguena,$^{13}$
S.~Allam,$^{14}$
F.~Andrade-Oliveira,$^{15}$
J.~Annis,$^{14}$
E.~Bertin,$^{16,17}$
S.~Bocquet,$^{18}$
D.~Brooks,$^{19}$
D.~L.~Burke,$^{20,21}$
A.~Carnero~Rosell,$^{22,13,23}$
M.~Carrasco~Kind,$^{24,25}$
J.~Carretero,$^{26}$
F.~J.~Castander,$^{7,8}$
M.~Costanzi,$^{27,28,29}$
M.~E.~S.~Pereira,$^{30}$
S.~Desai,$^{31}$
H.~T.~Diehl,$^{14}$
P.~Doel,$^{19}$
S.~Everett,$^{32}$
I.~Ferrero,$^{33}$
D.~Friedel,$^{24}$
J.~Frieman,$^{14,34}$
J.~Garc\'ia-Bellido,$^{35}$
M.~Gatti,$^{36}$
E.~Gaztanaga,$^{7,8}$
D.~Gruen,$^{18}$
J.~Gschwend,$^{13,37}$
G.~Gutierrez,$^{14}$
S.~R.~Hinton,$^{6}$
D.~L.~Hollowood,$^{38}$
K.~Honscheid,$^{39,40}$
D.~J.~James,$^{4}$
M.~March,$^{36}$
F.~Menanteau,$^{24,25}$
R.~Miquel,$^{41,26}$
R.~Morgan,$^{42}$
A.~Palmese,$^{43}$
F.~Paz-Chinch\'{o}n,$^{24,44}$
A.~Pieres,$^{13,37}$
A.~A.~Plazas~Malag\'on,$^{45}$
A.~K.~Romer,$^{46}$
E.~Sanchez,$^{47}$
V.~Scarpine,$^{14}$
I.~Sevilla-Noarbe,$^{47}$
M.~Soares-Santos,$^{15}$
E.~Suchyta,$^{48}$
G.~Tarle,$^{15}$
C.~To,$^{39}$
and T.~N.~Varga$^{49,50,51}$
\begin{center} (DES Collaboration) \end{center}
}}
\vspace{0.4cm}

\date{Accepted XXX. Received YYY; in original form ZZZ}

\pubyear{2022}

\begin{document}

\label{firstpage}
\pagerange{\pageref{firstpage}--\pageref{lastpage}}
\AddToShipoutPictureBG*{%
  \AtPageUpperLeft{%
    \hspace{0.75\paperwidth}%
    \raisebox{-4.5\baselineskip}{%
      \makebox[0pt][l]{\textnormal{FERMILAB-PUB-22-366-PPD}}
}}}%
\maketitle

\begin{abstract}
Type Ia supernovae (SNe Ia) are used as standardisable candles to measure cosmological distances, but differences remain in their corrected luminosities which display a magnitude step as a function of host galaxy properties such as stellar mass and rest-frame $U-R$ colour. Identifying the cause of these steps is key to cosmological analyses and provides insight into SN physics. Here we investigate the effects of SN progenitor ages on their light curve properties using a galaxy-based forward model that we compare to the Dark Energy Survey 5-year SN Ia sample. We trace SN Ia progenitors through time and draw their light-curve width parameters from a bimodal distribution according to their age. We find that an intrinsic luminosity difference between SNe of different ages cannot explain the observed trend between step size and SN colour. The data split by stellar mass are better reproduced by following recent work implementing a step in total-to-selective dust extinction ratio $(R_V)$ between low- and high-mass hosts, although an additional intrinsic luminosity step is still required to explain the data split by host galaxy $U-R$. Modelling the $R_V$ step as a function of galaxy age provides a better match overall. Additional age vs. luminosity steps marginally improve the match to the data, although most of the step is absorbed by the width vs. luminosity coefficient $\alpha$. Furthermore, we find no evidence that $\alpha$ varies with SN age.
\end{abstract}




\section{Introduction}

Empirical relationships among light-curve properties of type Ia supernovae (SNe Ia) are routinely used to standardize their peak brightnesses \citep{Pskovskii1977,Phillips1993,Riess1996,Tripp1998,Mandel2017} and facilitate their use as distance indicators in cosmological measurements \citep{Riess1998,Perlmutter1999,Sullivan2011,Betoule2014,Scolnic2018,DESCollaboration2018a,Brout2022}. This standardisation typically exploits the \lq faster--fainter\rq\ relation between peak brightness and SN light curve width, and the \lq bluer--brighter\rq\ relation between peak brightness and SN optical colour. A detailed astrophysical understanding of these relationships is not yet in place, but they are assumed to derive from physical processes intrinsic to the SN progenitors and their explosions (e.g., SN explosion mechanism and physics, or progenitor white dwarf mass, age or metallicity), or in the circumstellar medium (CSM) surrounding the explosion, or in the interstellar medium (ISM) between the SN and the observer (e.g., dust). Established correlations between the observed SN light-curve width and the properties of the SN host galaxy, with faster-declining SNe more readily located in more massive, less actively star-forming galaxies, presumably reflect some of this underlying astrophysics.

After standardising SN luminosities for SN light-curve width and SN colour, any remaining difference between inferred distances and those predicted by the cosmological model are termed \lq Hubble residuals\rq\,. These Hubble residuals are also observed to correlate with host galaxy properties, both those local to the SN and the integrated global properties. The most commonly used of these correlations is the between Hubble residual and host galaxy stellar mass \citep{Kelly2010,Lampeitl2010,Sullivan2010,Childress2013a}, which is typically represented by a step function in SN peak brightness of order $0.1$~mag at a host galaxy stellar mass of around $\log(M_*/\mathrm{M}_{\odot})=10$. A similar step in peak brightness is also observed when considering other host galaxy properties, such as specific star-formation rate \citep[sSFR;][]{Sullivan2010,Rigault2018}, gas phase metallicity \citep{Gallagher2005,Childress2013a}, rest-frame galaxy colour \citep{Roman2018,Kelsey2021}, and emission line equivalent widths \citep{Dixon2022}.
By assuming that the step is caused by two populations of SNe Ia that can be (imperfectly) probed by one of these  environmental tracers, \citet{Briday2021} showed that the size of the step is correlated with the level of population contamination across the step location of the given host galaxy tracer, with spectroscopically observed local sSFR and global stellar mass displaying the least contamination and largest steps. One interpretation of local sSFR being the best step tracer is that SNe Ia of different ages have different standardised luminosities, perhaps representing two (or more) SN Ia populations. Indeed, the case for multiple populations is strengthened by \citet{Nicolas2020} who show that the distribution of light-curve widths can be modelled as a combination of a young, slow-declining population and an old, fast-declining population.

Recent analyses have revealed a dichotomy in sizes of host galaxy -- Hubble residual steps when the SN samples are divided based on the colour of the SNe. In the 3-year spectroscopically-confirmed sample of SNe Ia from the Dark Energy Survey  \citep[DES3YR;][]{Brout2019}, redder SNe have large steps as a function of host stellar mass, while bluer SNe have negligible steps \citep{Kelsey2021}. This effect is also present in the larger, photometrically-classified DES5YR sample \citet[][hereafter K22]{Kelsey2022}, who also showed that the effect is present when considering host galaxy rest-frame $U-R$ colour in place of stellar mass. 

To explain the complex relationships between host galaxy Hubble residual steps and SN colour 
\citetalias{Brout2020} introduced a framework where the slope of the dust extinction law along the SN line-of-sight, as governed by the total-to-selective extinction parameter $R_V$, is different in low mass (mean $R_V = 2.75$) and high mass (mean $R_V = 1.5$) SN host galaxies. 
The effect of this model is that dustier SNe are redder and fainter, but the level of dimming is different for a SN with the same colour depending on the stellar mass of its host galaxy. 
Recent analysis by \citet{Popovic2021a} of the larger Pantheon+ SN Ia compilation \citep{Scolnic2021} confirms the results from \citetalias{Brout2020}, and shows that correcting SN Ia luminosities using this model may improve their use as cosmological probes. 
However, although the \citetalias{Brout2020} model reproduces the divergence of the stellar mass step as a function of SN colour, \citetalias{Kelsey2022} show that after correcting for the trend of Hubble residual versus SN colour when split by stellar mass, there remains a small residual step in host galaxy $U-R$, with SNe in redder galaxies appearing brighter after correction. As this $U-R$ colour is related to stellar population age, this suggests there may also be an age-based effect in addition to dust.

In this work, we build upon the model of \citetalias{Brout2020} to qualitatively investigate the relationships between host galaxy mass, $R_V$, galaxy colour, and age, with the aim of explaining the residual $U-R$ step observed by \citetalias{Kelsey2022} by expanding upon the multiple SN Ia age populations modelled by \citet{Nicolas2020}.

Most modern simulations of SN Ia data sets use the SuperNova Analysis (\textsc{snana};\citealt{Kessler2009}) software which constructs an accurate representation of how a large sample of SNe would appear in a survey, at the image level, by passing synthesised light curves through telescope responses and survey cadences. In this analysis we build a simplified toy model of the DES-SN survey by approximating many of the complex steps involved in \textsc{snana}, such as the light curve generation and fitting. This streamlining reduces computation time, allowing us to explore a broad parameter space while tracing individual stellar populations through a large simulation of galaxies. Results of this qualitative analysis will then be used to inform full scale simulations using \textsc{snana} in the future.

We base our simulation on a forward model of host galaxy properties, and test the effects of galaxy and SN age on the light curves, dust extinctions and HRs of SNe Ia. While most SN simulations use empirical relationships to assign SNe to host galaxies that have minimal information (usually stellar mass and SFR), here we use the stellar mass assembly models of \citet[][hereafter W21]{Wiseman2021} in order to create a library of host galaxies for which the stellar age distribution of the galaxy is known at any given look-back time. We then derive a SN progenitor age distribution for each galaxy in the library, allowing us to trace the age of each simulated SN. In Section \ref{sec:sims_hosts} we describe the galaxy simulation process and how we use stellar population synthesis models to translate star-formation histories into observables used in \citetalias{Kelsey2022}: stellar mass and rest-frame colour. In Section \ref{sec:sims_SNe} we describe various models that we use to sample light-curve parameters for the SNe, paying notable attention to the SN progenitor age and host stellar ages. Section \ref{sec:validation} presents a validation that the model approximates the observed SN parameter distributions and light-curve parameter -- host galaxy relations. In Section \ref{sec:results} we evaluate the models based on their ability to reproduce the Hubble residual trends with SN colour, and in Section \ref{sec:discussion} we discuss the interpretations and implications of the results.

Where appropriate we assume a flat $\Lambda$CDM cosmological model with $H_0 = 70~\mathrm{km~s}^{-1}~\mathrm{Mpc}^{-1}$ and $\Omega_{\mathrm{M}}=0.3$. Uncertainties are given as $1\sigma$ confidence intervals.
\begin{figure}

	\includegraphics[width=.5\textwidth]{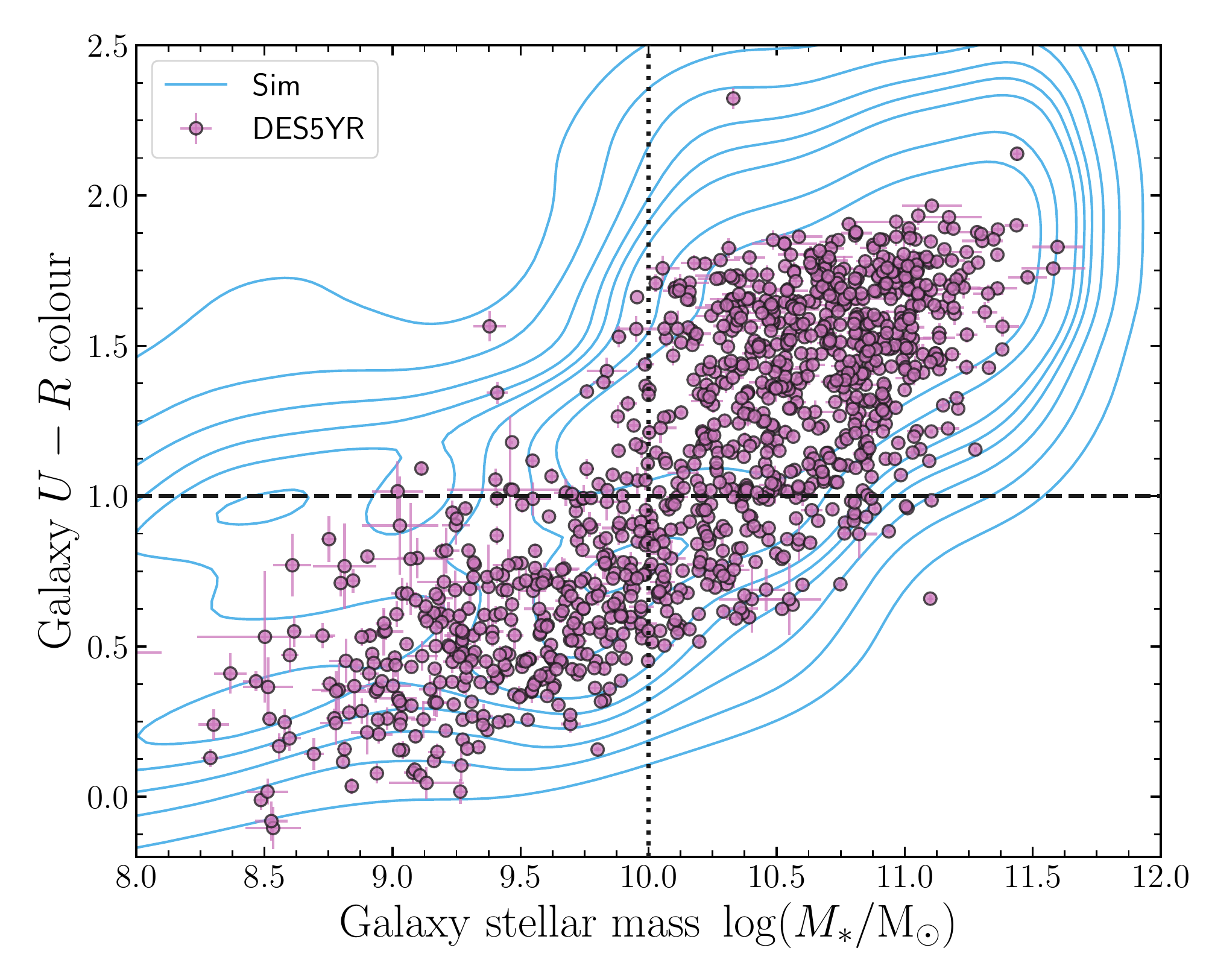}
    \caption{Rest frame galaxy $U-R$ colour versus stellar mass for the full simulated galaxy catalogue (blue contours) using the star-formation history models from \citetalias{Wiseman2021}, updated to include better quenching prescription. Magenta points are the observed stellar masses and colours from the DES5YR sample of SN Ia host galaxies (\citetalias{Kelsey2022}), and indicate where SN Ia hosts lie in this parameter space. The vertical dotted line indicates the $\log(M_*/\mathrm{M}_{\odot})=10$ location assumed for mass-related transitions throughout this work, while the horizontal dashed line is located at the corresponding point used to separate galaxy colours, $U-R =1$.} 
    \label{fig:mass_UR}
\end{figure}

\begin{figure*}

	\includegraphics[width=\textwidth]{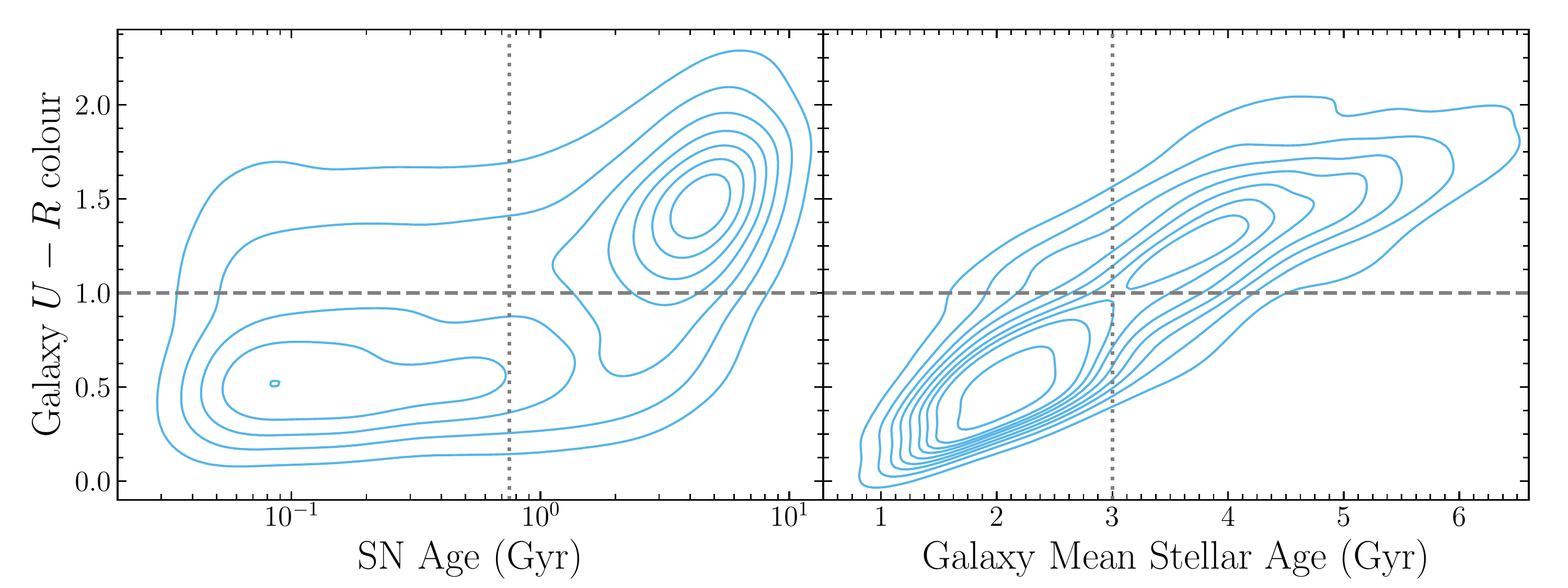}
    \caption{\textit{Left}: Galaxy rest frame $U-R$ colour vs SN progenitor age at the time of explosion for SNe in the simulation. Contours represent lines of equal relative density. The horizontal line represents the colour $U-R=1$ used to split observed populations, while the vertical line is the age used to split the young and old SN populations.
    \textit{Right}: As left, but galaxy rest frame $U-R$ colour vs galaxy mass weighted mean stellar age. The vertical line shows the 3~Gyr split used for distinguishing young and old galaxies.} 
    \label{fig:UR_Ages}
\end{figure*}

\section{Galaxy simulations}
\label{sec:sims_hosts}
The main aim of this work is to create a realistic simulation of a population of SNe Ia with an accurate representation of the links between the light-curve properties of the SNe and the global properties of their host galaxies. This library is then used to link the host galaxy properties to Hubble residuals. In this section, we outline the method used to simulate the SN Ia host galaxies and their properties. In Section \ref{sec:sims_SNe} we describe how we use the host galaxy library to simulate the SNe themselves.

We base our host galaxy simulation on the stellar mass assembly models of \citetalias{Wiseman2021}, which themselves are based on those of \citet{Childress2014}. These simple analytical models of galaxy evolution are derived empirically from observed relationships between stellar mass, SFR and redshift, and take into account mass loss and quenching. 

The ability of the models to reproduce observed stellar mass and SFR distributions is presented in \citet{Childress2014} and \citetalias{Wiseman2021}.
\subsection{Updated galaxy models}
\label{subsec:quenching}
In the simulations of \citetalias{Wiseman2021}, galaxies evolve following a deterministic set of equations, such that a simulated galaxy represents the average galaxy of a given stellar mass and redshift. While such a prescription is adequate for an overall modelling of SN rates as a function of mass and redshift, it does not result in an accurate \textit{distribution} of host galaxy star formation rates, which is important when considering host galaxy colours. To mitigate this issue, we update our prescription of galaxy quenching to be more stochastic. We also add bursts of star formation to the models. These two additions are detailed in turn below.

\subsubsection{Quenching}
Each galaxy track in the simulation is represented by $100$ implementations. At each time step of the simulation, each galaxy implementation is designated as either `non-quenched' or `quenched'. Quenched galaxies remain quenched, while non-quenched galaxies are subject to being quenched according to a quenching probability. This probability is a function of the stellar mass, similar to the average quenching penalty in the \citet{Childress2014} and \citetalias{Wiseman2021} implementations of the model:

\begin{equation}
    p_Q(M_*,z) = \frac{1}{2}\left[1 - \mathrm{erf}\left(\frac{\log(M_*) - \log(M_Q(z))}{\sigma_Q}\right)\right] - f_{\mathrm{Q}}\,,
\label{eq:pq}
\end{equation}
where $f_Q$ is the fraction of the $100$ implementations that have already quenched. In order to accurately reproduce the star formation main sequence and quenched fraction, we slightly adjust the reference quenching mass from \citetalias{Wiseman2021} by $0.2$~dex to:

\begin{equation}
    \log \left(M_Q(z)/\mathrm{M}_{\odot}\right) = 10.377 + 0.636z\,,
\end{equation}
and the transition scale $\sigma_Q$ from $1.1$ to $1.5$.

\subsubsection{Bursts of star formation}
A second update to the models of \citetalias{Wiseman2021} is the addition of random bursts of star formation to the models. We add instantaneous, delta-function bursts that form stars at a rate of half of the main sequence value for the mass and redshift of the galaxy. Bursts occur at a constant probability of 0.05 for every time step in the simulation.

\subsection{Simulating observable galaxy properties}

We build upon the model SFHs by ascribing observable properties to each of the simulated galaxies. In particular, we are interested in the observed and rest frame spectral energy distributions (SEDs), which are the observed data used to infer global galaxy properties such as stellar mass, SFR, and stellar population age. For the latter, we focus on the rest frame absolute magnitudes in Bessell following \citet{Roman2018}, \citet{Kelsey2021}, and \citetalias{Kelsey2022}. In the simulation, we link the model SFHs (with known stellar mass, SFR, and age distributions) to synthetic broad-band photometry using a similar method to that widely used in the inference of those observables: we convolve the SFHs with a grid of spectral templates of simple stellar populations (SSPs, Section \ref{subsec:libraries}) of different ages, summing up the individual SSP spectra according to the relative weights given by the SFH to produce a final composite spectrum $S_{\mathrm{gal}}$. The luminosities of the combined spectra are scaled by the stellar mass of the galaxy, resulting in an accurate representation of the intrinsic rest-frame galaxy spectra. We add nebular emission from ionised gas (Section \ref{subsec:nebular}), and apply a reddening due to interstellar dust via attenuation (Section \ref{subsec:gal_attenuation}). This grid of host galaxies acts as a host galaxy library from which we sample a realistic SN Ia host population in Section \ref{sec:sims_SNe}.

\subsubsection{Template libraries}
\label{subsec:libraries}

The choice of spectral templates affects the relationship between galaxy properties and observables. We use the common \citet[][hereafter BC03]{Bruzual2003} library, and for consistency with the SN host galaxy literature we generate SSPs with a \citet{Chabrier2003} initial mass function (IMF). The BC03 code draws upon the Padova 1994 evolutionary tracks \citep{Bertelli1994}. For simplicity, all SSPs are generated at solar metallicity ($Z_{\odot}$).

\subsubsection{Nebular emission}
\label{subsec:nebular}

In addition to the stellar continuum, our spectral templates incorporate nebular emission lines from \ion{H}{II} regions. The BC03 models do not incorporate nebular emission, and thus we add it separately following \citet{Boquien2019}. The line strengths are set relative to the ionizing photon flux, and are determined by the ratios at solar metallicity presented in \citet{Inoue2011}. The ionization parameter $U$ is set at $\log(U)=-2$.

\subsubsection{Dust attenuation}
\label{subsec:gal_attenuation}

To arrive at realistic simulation of galaxy spectra, we apply the effects of dust attenuation to the final scaled spectral templates. We add dust according to the attenuation law of \citet*{Cardelli1989}. For each galaxy there are 15 realisations with dust added in steps of equal size in $V$-band extinction $A_V$ in the range $0 \leq A_V \leq 1.5$~mag; the final value of $A_V$ for each host is chosen at a later stage along with the SN parameters, such that host has the same reddening $E(B-V)$ as the SN. To calculate the host extinction $A_V$ we follow the results of \citet{Salim2018} and use a total-to-selective extinction $R_V$ of $2.61$ for star-forming galaxies with $\log(M/\mathrm{M}_{\odot})< 9.5$, $2.99$ for  $9.5 \leq \log(M/\mathrm{M}_{\odot})< 10.5$, and $3.47$ for $\log(M/\mathrm{M}_{\odot})\geq 10.5$. For passive galaxies with $\log (\mathrm{sSFR})<-11$ we follow \citet{Salim2018} and assume the quiescent galaxy $R_V$ of 2.61.

\subsubsection{Synthetic galaxy photometry}
\label{subsubsec:sim_gal_properties}

We convert our simulated galaxy spectra into photometric observables by multiplying them with the transmission functions of desired passbands. This is possible with any filter at any wavelength -- the galaxy templates extend from the far ultraviolet to the far infrared -- but here we focus on optical wavelengths. 
For rest-frame galaxy photometry we calculate absolute magnitudes in the $UBVRI$ filters of \citet{Bessell1990} following \citetalias{Kelsey2022}, and in the observer frame we calculate apparent magnitudes in the DES $griz$ filters (e.g., $m_g^{\mathrm{host}}$, $m_r^{\mathrm{host}}$; \citealt{Flaugher2015})  for use in SN survey simulations \citep[e.g.,][]{Vincenzi2020}. 

The resulting library of simulated galaxy stellar masses and rest-frame colours are shown in Fig.~\ref{fig:mass_UR}, with DES SN Ia hosts shown to illustrate how the simulation covers the necessary stellar-mass--colour parameter space from which to accurately draw SN hosts.
\begin{figure}

	\includegraphics[width=.5\textwidth]{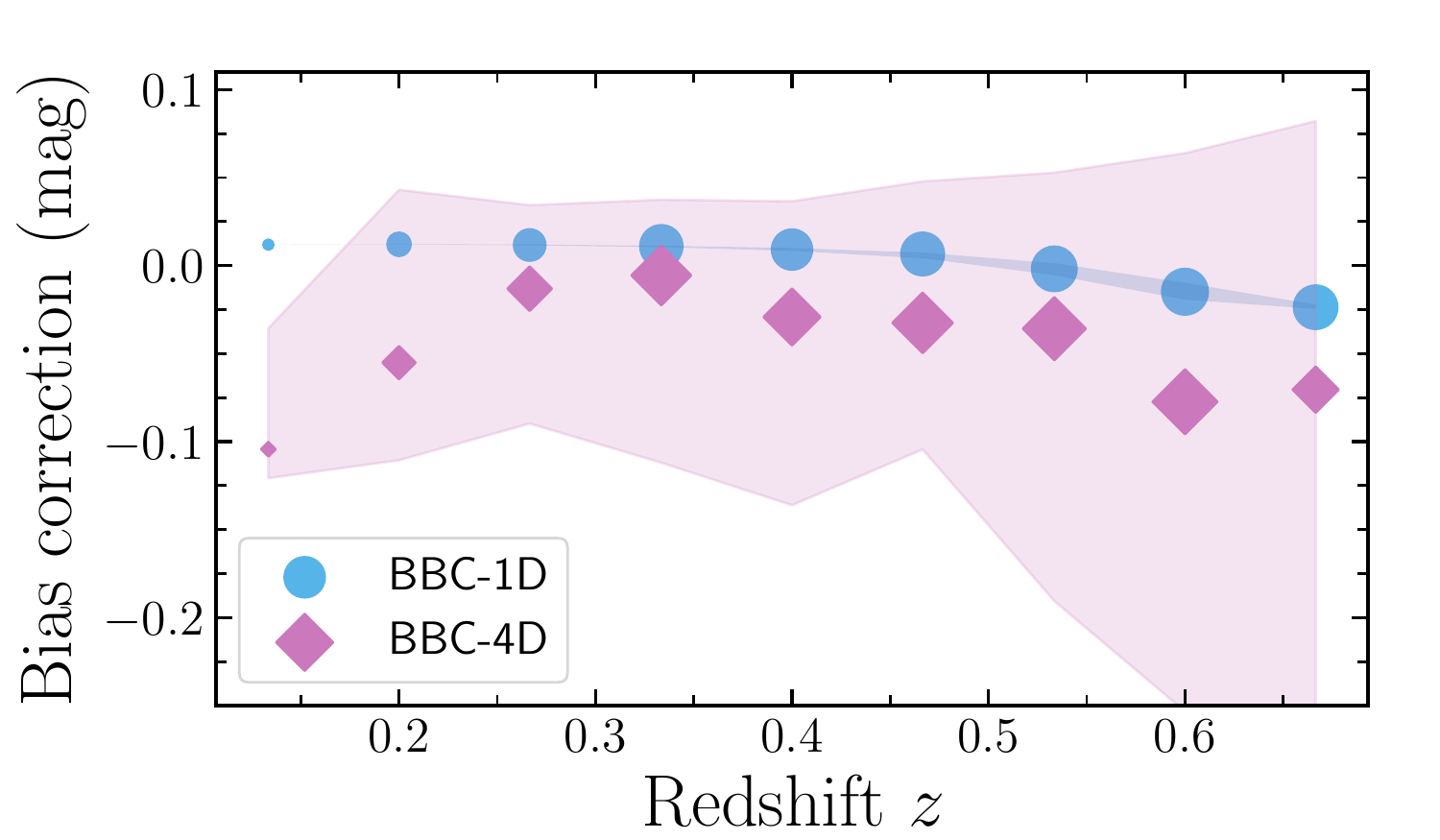}
    \caption{Average magnitude of a one-dimensional (BBC-1D; blue circles) and four-dimensional (BBC-BS20; magenta diamonds) bias correction as a function of redshift for a representative simulation. Points show the median in redshift bins; point sizes are proportional to the number of SNe in each redshift bin, while shaded regions show the 16th to 84th percentiles.  } 
    \label{fig:biascor_vs_z}
\end{figure}

\section{Supernovae}
\label{sec:sims_SNe}

We next describe how we simulate a sample of SNe Ia. This includes ensuring a realistic association with their host galaxies, accurate distributions of their light-curve parameters, and the principle effects introduced by observational noise. We discuss each in turn.

\subsection{SN host galaxy association}

We start with our simulated galaxy models from Section~\ref{sec:sims_hosts}, and take a flat distribution in stellar mass over the range $7<\log(M_*/\mathrm{M}_{\odot})<12$. We use a SN Ia delay-time distribution (DTD), which describes the
rate at which SNe Ia occur as a function of the delay time $\tau_{\mathrm{A}}$ since an episode of star formation, and convolve this DTD with each galaxy's SFH. In our simulations we use the DTD of \citetalias{Wiseman2021}, which takes a power-law form with index $-1.13$ and takes effect $40~$Myr after star formation. The result is a SN Ia rate per year in each model galaxy ($R_{\mathrm{G}}$). We then calculate a volumetric rate of SNe by multiplying $R_{\mathrm{G}}$ by a stellar-mass function ($\phi(M_*,z)$) at the redshift of the simulated galaxy. The combination $R_G\times\phi(M_*,z)$ is then used as a weighting to select a representative sample of SN Ia host galaxies in which SNe can be simulated. We use the redshift-dependent stellar-mass function from \textsc{zfourge} \citep{Tomczak2014}. 

The relationship between observed galaxy colour and SN and galaxy ages is shown in Fig. \ref{fig:UR_Ages}. There is a distinction between the host $U-R$ populations for young and old SNe, although there is a significant fraction of old SNe that exist in blue hosts. The relationship between galaxy colour and galaxy age is even clearer, with very little crossover between the old-and-red, and young-and-blue populations.

\subsection{Light curve properties}
\label{subsec:sim_lc}

For each simulated SN in a given host galaxy, we then assign \lq observed\rq\ light-curve properties by drawing randomly from statistical distributions. 
We use the standard SALT2 light-curve fitter \citep{Guy2007} framework of $x_1$ (light-curve width, or \lq stretch\rq ) and $c$ (SN peak rest-frame optical colour), together with a \citet{Tripp1998}-like standardisation relation that relates the inferred distance modulus $\mu_\mathrm{obs}$ to SN Ia light-curve parameters. In most current SN Ia analyses, this takes the form
\begin{equation}
    \mu_\mathrm{obs}=m_B - M_B + \alpha x_1 - \beta c + \mu_\mathrm{bias}\,,
    \label{eqn:tripp}
\end{equation}
where $M_B$ is the peak SN absolute magnitude in the rest-frame $B$ band, $m_B$ is the peak rest-frame apparent magnitude, the $\alpha$ and $\beta$ coefficients parametrize the stretch--luminosity and colour--luminosity relationships, and the $\mu_\mathrm{bias}$ term is a correction for biases introduced by survey selection effects. Hubble residuals, $\mu_{\mathrm{res}}$, can then be calculated using 
\begin{equation}
    \mu_{\mathrm{res}} = \mu_{\mathrm{obs}} - \mu_{\mathrm{model}}\,,
\end{equation}
where $\mu_{\mathrm{model}}$ depends on the cosmological parameters, which in this analysis we hold fixed at our fiducial model. 

This simple framework assumes all colour variation in SNe Ia can be captured in a simple \lq $\beta c$\rq\ term. However, we are explicitly interested in this analysis in the effect of dust extinction and reddening on SNe Ia, and thus we also assign each simulated SN a colour excess (i.e., $E(B-V)$) due to dust reddening that we denote $E_\mathrm{SN}$, and a total-to-selective extinction parameter $R_V$. We next outline the methods used to draw the samples for each light-curve and dust parameter.

\subsubsection{SN Ia absolute magnitude}
\label{subsubsec:sim_MB}

We simulate SNe using a fixed peak $B$-band absolute magnitude of $M_B=-19.325$~mag. We allow for the possibility of host galaxy \lq steps\rq\ of size $\gamma$, where SNe on either side of a threshold in a given host galaxy property $X$ (such as stellar mass) have a factor of $\Gamma \times \gamma$ added to their simulated absolute magnitude, where $\Gamma =+0.5$ for hosts below the threshold and $\Gamma =-0.5$ for hosts above the threshold $X_{\mathrm{split}}$. Thus,
\begin{equation}
    M_B = -19.325 + \Gamma_X \times \gamma_X ~\mathrm{mag}\,,
    \label{eq:mb_step}
\end{equation}
with
\begin{equation}
 \Gamma_X = \left\{
    \begin{array}{@{}rc@{}}
    0.5, & X < X_{\mathrm{split}} \\
    -0.5 , & X \geq X_{\mathrm{split}} \\
    \end{array}\right. \,,
        \label{eq:Gamma}
\end{equation} 
where $X_{\mathrm{step}}$ is the location of the step in the given host parameter.

\subsubsection{SN $x_1$}
\label{subsubsec:sim_x1}

We include two prescriptions for the SALT2 $x_1$ parameter. The first parametrises the $x_1$ distribution as an asymmetric Gaussian, while the second assumes two Gaussians dependent on the age of the SN progenitor's stellar population. 

The first $x_1$ population model is used in \citetalias{Brout2020}, which in turn follows \citet{Scolnic2016}, where $x_1$ is drawn from an asymmetric Gaussian model defined by parameters $\sigma_-, \sigma_+, \overline{x_1}$ \footnote{We use the notation $\overline{x_1}$ instead of the conventional $\mu_{x_1}$ to avoid confusion with the distance modulus $\mu$, and maintain this notation for all parameter means.}, giving a probability distribution
\begin{equation}
 P(x_1) = \left\{
    \begin{array}{@{}cc@{}}
    e^{[-(x_1-\overline{x_1})^2/2\sigma^2_-]} , & x_1<\overline{x_1}\\
   e^{[-(x_1-\overline{x_1})^2/2\sigma^2_+]} , &  x_1>\overline{x_1} \\
    \end{array}\right. \,.
        \label{eq:x1_asymm}
\end{equation} 

The second $x_1$ model builds upon the evolving double-Gaussian of \citet{Nicolas2020} with the $x_1$ probability distribution represented as a Gaussian mixture model -- a combination of two Gaussian distributions, one for a \lq young\rq\ SN population and one an \lq old\rq\ population. In the \citet{Nicolas2020} model, environmental age is traced by the local specific SFR (lsSFR). SNe in young (high lsSFR) environments are assumed to belong solely the old $x_1$ population, while SNe in old (low lsSFR) environments can be drawn from either the old or young $x_1$ modes, with the relative weight of each determined by a mixing probability $a_{x_1}$. Since the fraction of old and young SNe varies with redshift, so does the $x_1$ distribution (see Fig.~8 in  \citealt{Nicolas2020}).

Instead of using lsSFR as an observational indicator of progenitor age, our simulations allow us to track the progenitors directly. For each galaxy, the likelihood that a SN exploding at time $t_0$ comes from a progenitor of age $\tau_{\mathrm{A}}$ is
\begin{equation}
    P(t_0,\tau_{\mathrm{A}}) = \psi(t_0-\tau_{\mathrm{A}})\Phi(\tau_{\mathrm{A}})\,,
    \label{eq:prog_age_dist}
\end{equation}
where $\psi(t_0-\tau_{\mathrm{A}})$ is the SFH and $\Phi(\tau_{\mathrm{A}})$ the DTD. We sample SN age $\tau_{\mathrm{A}}$ from this distribution, and subsequently derive $x_1$ from the age by adapting the \citet{Nicolas2020} prescription
\begin{equation}
    x_1 \sim \left\{
    \begin{array}{@{}cc@{}}
    \mathcal{N}(\overline{x_{1,1}},\sigma_{x_1,1}), &  \tau_{\mathrm{A}}< \tau_{\mathrm{A, thresh}} \\[5pt]
   a_{x_1} \times\mathcal{N}(\overline{x_{1,1}},\sigma_{x_1,1}) +\\ (1-a_{x_1})\times\mathcal{N}(\overline{x_{1,2}},\sigma_{x_1,2}) , & \tau_{\mathrm{A}}> \tau_{\mathrm{A, thresh}}\\
  
    \end{array}\right.
\end{equation}
where $\overline{x_{1,[1,2]}}$ and $\sigma_{x_1,[1,2]}$ are the mean and standard deviation of the young and old Gaussians respectively, $a_{x_1}$ is the mixing coefficient which corresponds to the probability that the progenitor belongs to the young $x_1$ mode, and $\tau_{\mathrm{A,~thresh}}$ is the age which separates young and old timescales.

\begin{table}
	\centering
	\caption{Outline of the different SN Ia models used in this analysis.}
	\label{tab:models}
	\begin{tabular}{lcll} 
		\hline
		Model name & SN $x_1$   & SN $c$ & $R_V$ \\
		&distribution&distribution&distribution\\
		\hline
		BS21 & SK16  & Gauss + exp (mass) &Step (stellar mass) \\
	    Age $R_V$ & N21 &  Gauss + exp (age)  &Step (galaxy age)\\
	    Age $R_V$ Linear  &N21 & Gauss + exp (age)  & Linear (WD age)\\
	    
		\hline
	\end{tabular}
\end{table}
\begin{table*}
	\centering
	\caption{SN population parameters for the age-based models. Explanation of parameters can be found in Section~\ref{subsec:sim_lc}.
	}
	\label{tab:age_SN_params}
	\begin{tabular}{cccccccccc} 
		\hline
		 $\overline{x_{1, 1}}$  & $\sigma_{x_1, 1}$  &  $\overline{x_{1, 2}}$  & $\sigma_{x_1, 2}$  & $a_{x_1}$ & $\tau_{\mathrm{A,~thresh}}$& $\overline{c_{\mathrm{int}}}$  & $\sigma_{c_\mathrm{int}}$ & $\tau_{E,1}$  & $\tau_{E,2}$\\
		\hline
		0.22 & 0.61 & $-1.22$ & 0.56 & 0.38& 0.75~Gyr & $-0.075$ & 0.042 & 0.135 & 0.135\\
	\end{tabular}
\end{table*}

\subsubsection{SN colour}
\label{subsubsec:sim_c}

We follow the method of \citetalias{Brout2020}, which itself builds on \citet{Jha2007} and \citet{Mandel2011,Mandel2017}. These model the observed SN Ia colour distribution as a combination of an intrinsic colour $c_{\mathrm{int}}$ and the colour excess due to dust reddening $E_{\mathrm{SN}}$.
$c_{\mathrm{int}}$ is normally distributed
\begin{equation}
    c_{\mathrm{int}}\sim \mathcal{N}(\overline{c_{\mathrm{int}}},\sigma_{c_{\mathrm{int}}})\,,
\end{equation}

and $E_{\mathrm{SN}}$ is drawn from an exponential distribution following \citetalias{Brout2020}
\begin{equation}
    P(E_{\mathrm{SN}}) = \frac{1}{\tau_E}e^{-E_{\mathrm{SN}}/\tau_E}\,,
\end{equation}
where $\tau_E$ is the mean reddening of the SN population.

\subsubsection{Extinction law $R_V$}
\label{subsubsec:sim_Rv}

Related to the SN dust reddening $E_{\mathrm{SN}}$ is the slope of the reddening law $R_V$, such that the  absolute $V$-band extinction $A_V$ is
\begin{equation}
    A_V = R_V E_\mathrm{SN}\,,
    \label{eq:Av}
\end{equation}
and for the $B$-band magnitudes typically used in SN Ia analyses,
\begin{equation}
    A_B = R_B E_\mathrm{SN} \equiv (R_V+1) E_\mathrm{SN}\,.
    \label{eq:Ab}
\end{equation}

We investigate three models for how $R_V$ varies among SNe. These are inspired by the model of \citetalias{Brout2020}, where $R_V$ follows a step with stellar mass, and motivated by the evidence that the $x_1$ distribution is 
related to SN or host galaxy age. Our $R_V$ that depend on some host galaxy property $Y$ are:
\begin{enumerate}
    
    \item BS21 ($Y = M_*$): $R_V$ is drawn from a Gaussian distribution:
    \begin{equation}
        R_V\sim \mathcal{N}(\overline{R_V},\sigma_{R_V})\,,
    \end{equation}
    with $\overline{R_V}$ and $\sigma_{R_V}$ taking on different values for host galaxies on either side of $Y_{\mathrm{split}}$, a split point in stellar mass;
    
    \item Age $R_V$ ($Y = \tau_{\mathrm{G}}$): The same model as BS21, but using mass-weighted mean stellar age of the host galaxy, $\tau_{\mathrm{G}}$, instead of host stellar mass, as the parameter to split $R_V$. $\tau_{\mathrm{G}}$ is linked to both the observed stellar mass and the $U-R$ colour;
    
    \item Age $R_V$ Linear ($Y = \tau_{\mathrm{A}}$): $\overline{R_V}$ follows a linear relationship with progenitor age $\tau_{A}$. We parametrise the relationship by fixing the value of $\overline{R_V}$ for very young ($t_1$) and very old ($t_2$) progenitors as
    \begin{equation}
        \overline{R_V} = R^{t_1}_V + \left(\tau_{\mathrm{A}}-t_1\right)\frac{R^{t_2}_V - R^{t_1}_V}{t_2 - t_1}\,,
    \end{equation}
    and the standard deviation $\sigma_{R_V}$ following a similar relationship in order to encapsulate having more diverse dust in younger systems:
     \begin{equation}
        \sigma_{R_V} = \sigma^{t_1}_{R_V} + \left(\tau_{\mathrm{A}}-t_1\right)\frac{\sigma^{t_2}_{R_V} - \sigma^{t_1}_{R_V}}{t_2 - t_1}\,.
    \end{equation}
\end{enumerate}

\subsubsection{Mean reddening $\tau_E$}
\label{subsubsec:sim_tauE}
For setting the mean reddening $\tau_E$ we follow \citetalias{Brout2020}, fixing the value separately in low- and high-mass galaxies. We also implement a slight variation where the split is defined on the mass-weighted mean stellar age of the galaxy.

\subsubsection{Colour-luminosity coefficient $\beta$}
\label{subsubsec:sim_beta}
As per \citetalias{Brout2020} we do not assume a universal  relationship between SN peak brightness and $c_{\mathrm{int}}$. Instead we allow the coefficient relating them, denoted $\beta_{\mathrm{sim}}$, to vary between SNe, and draw it from a Gaussian distribution:
\begin{equation}
    \beta_{\mathrm{sim}} \sim \mathcal{N}(\overline{\beta},\sigma_{\beta})\,,
\end{equation}
with mean $\overline{\beta}$ and standard deviation $\sigma_{\beta}$. In this model $\beta_\mathrm{sim}$ is independent of any host galaxy, progenitor, or dust property.

\subsubsection{Stretch-luminosity coefficient $\alpha$}
\label{subsubsec:sim_alpha}
For the baseline model we adopt a value of $0.15$ for the stretch-luminosity coefficient $\alpha_{\mathrm{sim}}$. In cases where a mass or age luminosity step are introduced, the measured value of $\alpha$ can be different to the input value $\alpha_{\mathrm{sim}}$. In such cases we vary $\alpha_{\mathrm{sim}}$ such that the measured value matches that found by \citetalias{Kelsey2022}. We further investigate the effects of $\alpha$ and its importance in age-based models in Section \ref{subsubsec:alpha}.

\begin{figure*}
	\includegraphics[width=\textwidth]{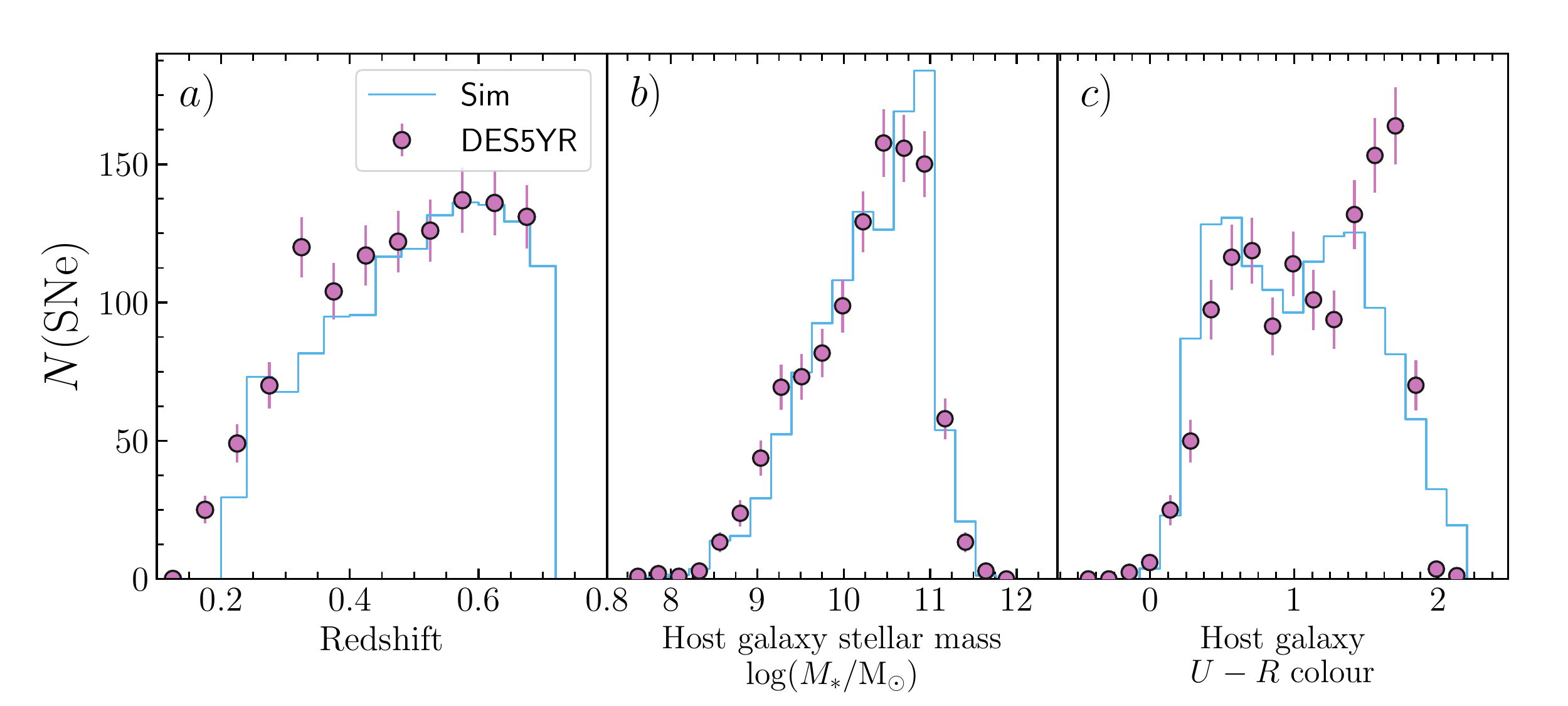}
    \caption{Population distributions of observed DES5YR SNe Ia compared to SNe Ia samples simulated in this work, in redshift, host galaxy stellar mass, and host galaxy rest-frame $U$-$R$ colour. Counts from the simulation have been scaled to match the total number of DES5YR SNe.} 
    \label{fig:dists}
\end{figure*}
\begin{figure*}

	\includegraphics[width=\textwidth]{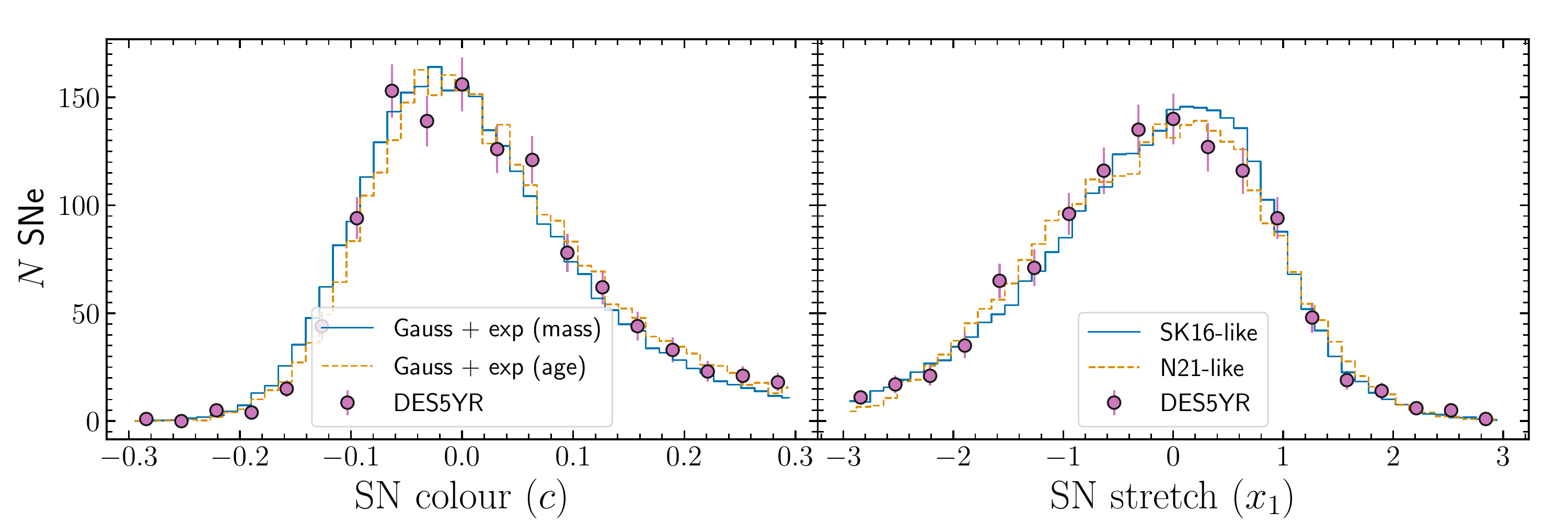}
    \caption{\textit{Left}: distribution of SN SALT2 colour $c$ for DES5YR and simulations drawing $c$ from an intrinsic Gaussian and an exponential dust distribution. \textit{Right}: distribution of SN SALT2 stretch $x_1$ for DES5YR and simulations drawing from an asymmetric Gaussian (yellow dashed) as per \citet{Scolnic2016} and a double Gaussian based on SN progenitor age (red solid) as per \citetalias{Nicolas2020}.} 
    \label{fig:SN_lc_samples}
\end{figure*}
\begin{figure*}

	\includegraphics[width=\textwidth]{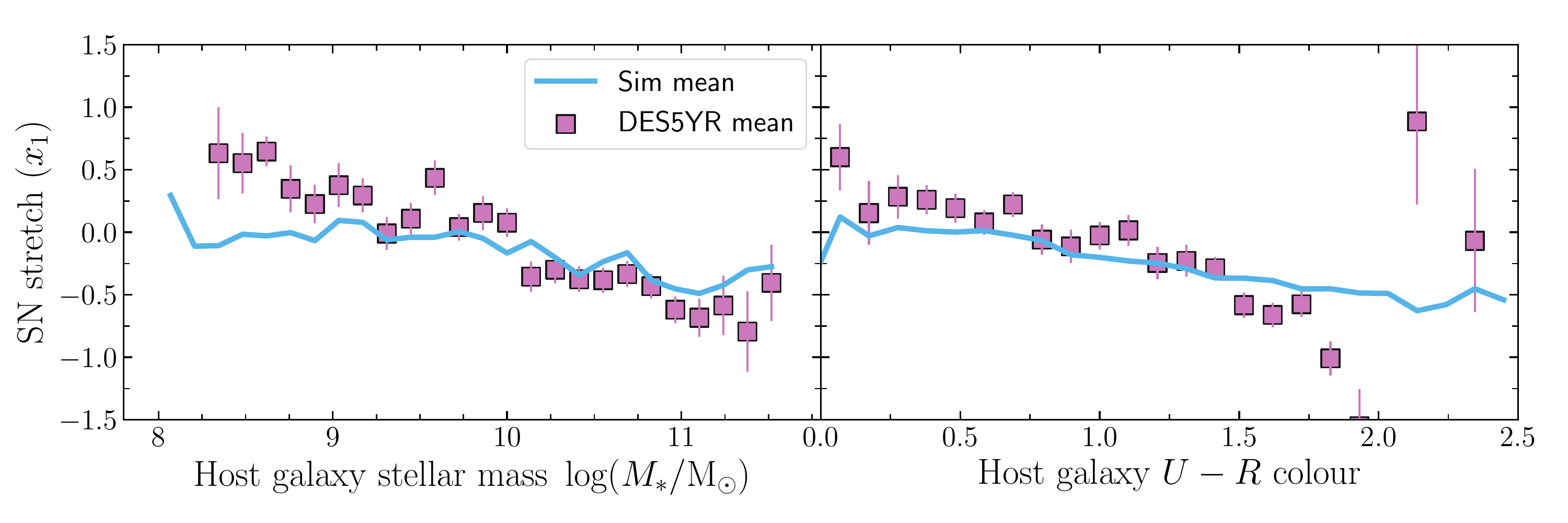}
    \caption{\textit{Left}: distribution of SN $x_1$ as a function of host galaxy stellar mass for DES5YR data and simulations using the \citetalias{Nicolas2020} double Gaussian $x_1$ model based on SN progenitor age; \textit{Right}: as left, but for host galaxy $U-R$ colour.}
    \label{fig:x1_vs_hosts}
\end{figure*}

\begin{figure*}

	\includegraphics[width=\textwidth]{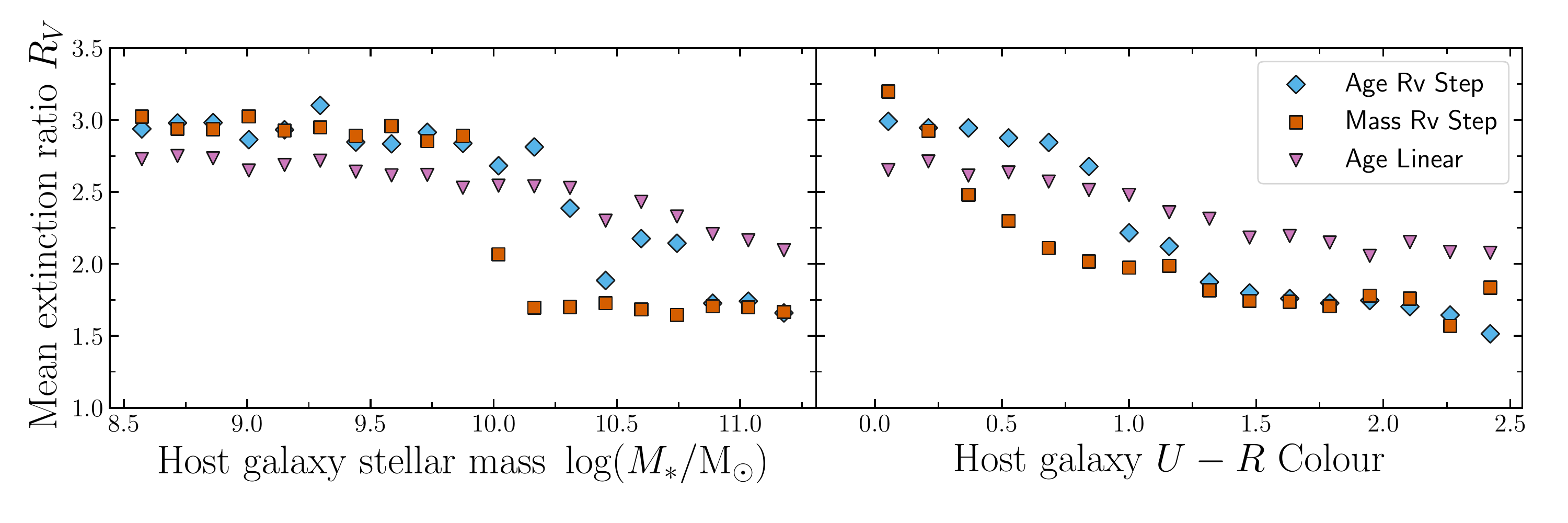}
    \caption{\textit{Left}: Relationship between total-to-selective dust extinction $R_V$ and host galaxy stellar mass for the three models used in this work. \textit{Right}: As left, but shown against host galaxy $U-R$ colour. } 
    \label{fig:Rv_v_gals}
\end{figure*}
\subsection{Observables}
\label{subsec:sim_mB}

Given a set of simulated SN Ia light curve parameters, the synthetic $m_B$ is then calculated by analogy to equation~\ref{eqn:tripp}, modified to explicitly distinguish dust from SN colour variations, i.e.,
\begin{equation}
    m_B = M_B + \mu(z)_{\mathrm{sim}} - \alpha_{\mathrm{sim}} x_1 +\beta_{\mathrm{sim}} c_\mathrm{int} + E_{\mathrm{SN}}(R_V +1) +  \Delta m_B\,,
\label{eq:tripp_bs20}
\end{equation}
where $\mu(z)_{\mathrm{sim}}$ is the distance modulus at the simulated redshift $z$ given by the reference cosmology. 

Even following the standardisation of SNe Ia there remains additional scatter in their observed $M_B$. We describe this in our simulated SNe using the parameter $\sigma_{\mathrm{int}}$. We draw an offset ($\Delta M_B$) to $M_B$ from this scatter according to the distribution:
\begin{equation}
    \Delta M_B \sim \mathcal{N}\left(0,\sigma_{\mathrm{int}}\right)\,.
    \label{eq:sigma_int}
\end{equation}
The value of $\sigma_{\mathrm{int}}$ is tuned so that in combination with the rest of the model the overall unexplained scatter in the Hubble residuals matches that of the data, which is of the order $\sim0.1~$mag.

\subsection{Application to DES5YR data}
The models introduced above are generic and intended to approximate the intrinsic universal properties of SNe Ia. Here we describe the data set against which we compare the simulations, and the survey-specific additions to the simulation that are necessary in order to replicate the selection effects of the data set.

We use an identical sample to that presented in \citetalias{Kelsey2022}: namely a photometrically classified SN Ia sample using five years of DES observations (DES5YR), as described in \citet{Vincenzi2020} and \citet{Moller2022}, with a redshift limit of $z\leq 0.7$. Despite the redshift cut, the sample is magnitude-limited, and thus there are selection effects which we must approximate, which we discuss below.

\subsubsection{Host galaxy spectroscopic efficiency}
\label{subsubsec:spec_eff}

To be included in the sample, DES SNe Ia require a spectroscopic redshift of the host galaxy, and the efficiency of obtaining one in the spectroscopic follow-up campaigns is a strong function of host galaxy apparent magnitude, $m_r^{\mathrm{host}}$. \citet{Vincenzi2020} presented the DES SN spectroscopic-redshift efficiency function $\epsilon_{z_{\mathrm{spec}}}(m_r^{\mathrm{host}})$, which we apply to our simulated SN hosts. In the simulation, we assign a probability of successfully obtaining a spectroscopic redshift for each galaxy at random from a Bernoulli distribution with success probability $p=\epsilon_{z_{\mathrm{spec}}}(m_r^{\mathrm{host}})$. In this work we do not model further survey efficiencies \citep[e.g., SN detection efficiency;][]{Brout2019a,Kessler2019}. Therefore, instead of simulating a population of SNe that follows the true intrinsic volumetric rate as a function of redshift, we sample redshift from an ad-hoc probability distribution that matches the observed DES5YR redshift distribution. We find that sampling with a probability that increases directly proportional to redshift (up until $z=0.7$) matches the data well. We validate this $z$ distribution in Section \ref{sec:validation}.

\subsubsection{Supernova uncertainties}
Without fully simulating SN light curves and their subsequent observation by the survey, we must approximate the uncertainties on our SN light curve parameters.
We simulate uncertainties $\sigma_{m_B}$ on $m_B$ by approximating the relationship between $\sigma_{m_{B},\mathrm{obs}}$ and $m_{B,\mathrm{obs}}$ in the DES5YR data. The uncertainty on $x_1$ ($\sigma_{x_1}$) is then simulated according to $\sigma_{m_B}$ following a linear least-squares fit to the observed DES5YR $\sigma_{m_B}$ and $\sigma_{x_1}$. The same procedure is followed for the colour uncertainty, $\sigma_c$.
A full description of the uncertainties is presented in Appendix~\ref{sec:app_errors}.

\subsubsection{Bias corrections}
\label{subsubsec:sim_HR}

With simulated brightnesses and light curve parameters, their respective uncertainties, as well as a redshift distribution that all match the data, the simulation mimics the selection effects and observational biases inherent in the data. We therefore calculate the distance moduli to each simulated SN Ia using an identical method to \citetalias{Kelsey2022} using the BEAMS with Bias Corrections \citep[BBC;][]{Kessler2017} framework. BBC follows the method of \citet{Marriner2011}, a $\chi^2$ minimization of Hubble residuals binned in redshift. The BEAMS part of BBC is not used, as we do not deal with contamination from core-collapse SNe and all simulated SNe are assigned as SNe Ia with probability $1.0$.
In BBC, we fit for Eq. \ref{eqn:tripp}. We do not fit for the additional host step terms $\gamma_X$, as we investigate how these terms evolve as a function of SN colour in this work.

The simplest form of bias correction ($\mu_\mathrm{bias}$; equation~\ref{eqn:tripp}) implemented in BBC is dependent on redshift only: a one-dimensional correction (\lq BBC-1D\rq). However, \citet{Scolnic2016} showed that the asymmetric nature of the $c$ and $x_1$ distributions leads to distance biases as a function of those parameters, which must be corrected as a function of $\alpha$, $x_1$,  $\beta$, and $c$ in addition to $z$ (a \lq BBC-5D\rq\ bias correction). In implementations of such 5D corrections, the treatment of relationships between $x_1$ and host stellar mass must also be carefully accounted for to avoid introducing subtle biases \citep{Smith2020}. 
\citet{Popovic2021} introduced a further two dimensions to account for such correlations between light-curves and host galaxies (\lq BBC-7D\rq), as well as a method that assumes the \citetalias{Brout2020} colour model (\lq BBC-BS20\rq).

In this work we use the 1D (redshift only) bias correction to validate and compare models, and also compare our best models to the data with the BBC-BS20 corrections. We include a $x_1$-host mass correlation in the bias correction simulation as per \citet{Vincenzi2020}, and for BBC-BS20 we use the best-fit $R_V$ values from \citetalias{Brout2020} in the bias correction simulation. Both approaches require that the simulation adequately includes the same observational biases and selection effects that have been introduced onto the data, which although not guaranteed has been well validated in the following section. The average and range of bias correction magnitudes for both cases are shown in Fig. \ref{fig:biascor_vs_z} and show that, in particular for the BBC-1D case, the corrections are small compared to the step size in the redshift range considered.

\subsection{Baseline model}

For the validation of the model (Section \ref{sec:validation}), we introduce baseline simulations on top of which the dust and luminosity step parameters are subsequently varied (Section \ref{sec:results}). For the \citetalias{Brout2020} model we fix the $x_1$ and $c$ population parameters similarly to those used in \citetalias{Brout2020}. We use the $x_{1,\mathrm{sim}}$ parameters from \citet{Scolnic2016}, while for $c_{\mathrm{sim}}$ we use the values in Table \ref{tab:age_SN_params}. The parameters for both $x_{1,\mathrm{sim}}$ and $c_{\mathrm{sim}}$ used in our baseline age-based model are also presented in Table \ref{tab:age_SN_params}. Dust parameters $R_V$ are held constant at $2.75$, although $R_V$ is not important for the validation presented below. The width-luminosity coefficient is fixed at $\alpha_{\mathrm{sim}}=0.15$ and the colour-luminosity coefficient has mean $\overline{\beta}=1.98$ and standard deviation $\sigma_{\beta}=0.35$

\section{Validation}
\label{sec:validation}

Before presenting the results, we describe the metrics used to validate the simulations introduced in Sections \ref{sec:sims_hosts} and \ref{sec:sims_SNe}. We validate our model by comparing it to the DES5YR SN Ia data presented in \citetalias{Kelsey2022}. A set of qualitative and quantitative validations are outlined as follows. In the following sections, we define the chi-squared statistic on a histogram with both the data and simulation in equivalent bins:
\begin{equation}
    \chi^2 = \sum_i \frac{\left(N_i^{\mathrm{data}} - N_i^{\mathrm{sim}}\right)^2}{e_i^2}\,,
\end{equation}
where $N_i^{\mathrm{data}}$ is the count of the observed data in the $i$th bin, $N_i^{\mathrm{sim}}$ is the equivalent for the simulation, and $e_i^2$ the error on the data approximated by $\sqrt{N_i^{\mathrm{data}}}$. While this estimate of the error does not take into account uncertainty on the measurements themselves, doing so would require a further level of resampling that would render the procedure computationally prohibitive. 
For goodness of fit and model comparison, we also employ the reduced chi-squared:
\begin{equation}
    \chi^2_{\nu} =\chi^2/N_{\mathrm{bins}}\,.
\end{equation}

\subsection{Host galaxies}
\label{subsec:res_valid_popns}

We validate the host galaxy simulations by investigating the one-dimensional distributions of redshift, stellar mass ($M_*$), and rest-frame $U-R$ (Fig.~\ref{fig:dists}). The overall trend of the redshift distribution is recovered by construction, as outlined in Section~\ref{subsubsec:spec_eff}. The distribution of observed $M_*$ is well-replicated. The observed distribution of $U-R$ is bimodal, and this general behaviour is recovered by the model, although the peak of red galaxies in the simulation lies around $U-R \sim1.5$, compared to the data which peak closer to $1.8$. This difference could be due to a number of reasons, such as the simplistic treatment of metallicity and an incomplete implementation of galaxy quenching. While not perfect, we proceed satisfied that both the observed and simulated distributions can be separated at $U-R$=1. Since the Hubble residual steps tested using this split point, differing shape of the red peak of the distribution should not affect the results significantly.

We further validate this argument by comparing the observed and simulated correlations between $M_*$ and $U-R$ (Fig.~\ref{fig:mass_UR}): the simulations accurately replicate the positive trend between $U-R$ and $M_*$, as well as the characteristic step from majority blue to majority red galaxies as their mass increases beyond $\log(M_*/\mathrm{M}_{\odot}) = 10$. This qualitative validation reinforces trust that our galaxy evolution model reproduces, on a general level, the complex relationships between galaxy stellar mass, age, and colour, and that the DTD--SFH convolution selects a representative sample of SN Ia hosts from the global galaxy population.

\subsection{Light curve parameters}
\label{subsec:res_valid_lcs}

We next validate the simulated distributions of SN $x_1$ and $c$, and their relationships with host galaxy attributes $M_*$ and $U-R$. The simulated distributions are shown in Fig.~\ref{fig:SN_lc_samples}, and the parameters used for the baseline age-based model are presented in Table~\ref{tab:age_SN_params}.

The $c$ distribution is well-matched by the \lq Gaussian plus dust reddening\rq\ model. We find the data are described well ($\chi^2_{\nu} = 0.95$) using similar parameters to \citetalias{Brout2020} for both the intrinsic Gaussian and exponential tail, with the mean of the colour distribution shifting marginally bluer. We find a single value of $\tau_E =0.135$ describes the data well, which implies there is no strong difference in the mean reddening for SNe in different galaxies, whether they are split by their stellar mass or their age. 

Both the asymmetric Gaussian model and Gaussian mixture model provide good fits to the $x_1$ distribution, with the mixture model offering a slightly improved fit around the peak. The best-matching parameters for the mixture model are similar to those from \citetalias{Nicolas2020}, with the only significant difference being the mixture coefficient $a_{x_1}$, for which \citetalias{Nicolas2020} report $\sim0.5$ while we find $0.38$. In our model, the value of $a_{x_1}$ is degenerate with the choice of age threshold $\tau_A$, an investigation into which is left for future work. There is a degree of overfitting, with the best fit model having $\chi^2_{\nu} = 0.65$, due to the large uncertainties on measured $x_1$ in the SN Ia data.

The relationship between $x_1$ and host galaxy properties is shown in Fig.~\ref{fig:x1_vs_hosts}. The \citetalias{Nicolas2020} model implemented in our simulations reproduces the general trends in the data, but does not achieve the same strength of relation, despite the one-dimensional $x_1$ distribution being adequately replicated (Fig.~\ref{fig:SN_lc_samples}). The likely cause is the inconsistency between the simulated and observed $U-R$ distribution, due to the under-representation of passive/red galaxies in the simulation. The underprediction of the slope of the $x_1$ versus $M_*$ relation has consequences in the recovery of the $\alpha$ parameter \citep{Dixon2021,Rose2021}.

\begin{figure*}

	\includegraphics[width=\textwidth]{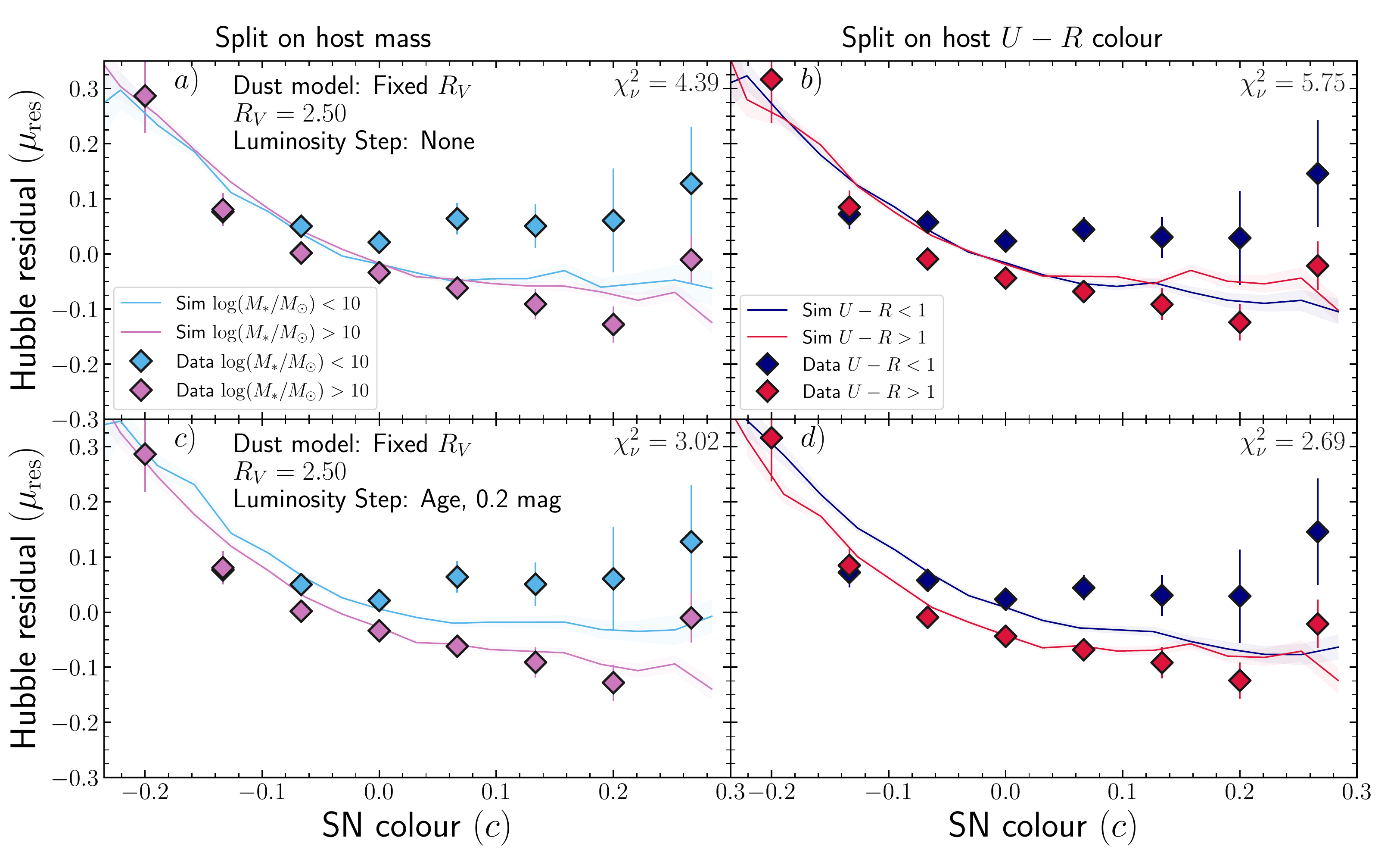}
    \caption{SN Ia Hubble residual versus SN colour, with the DES5YR SN Ia sample shown as the data points and the simulations shown as the solid lines. Left hand panels show SNe split by their host stellar mass at split at $\log(M/M_*)=10$, and right hand panels show SNe split by their host colour at  $U-R=1$.
    Panel \textit{a)} shows simulations from the baseline model with $R_V$ fixed across all hosts and no intrinsic luminosity step. Hubble residuals for data and simulations have been measured with BBC-1D. \textit{b)}: as the left panel, but for the data and simulations split at a host galaxy rest-frame $U-R$ colour of $U-R=1$. The $\chi^2_{\nu}$ statistic for both comparisons is shown.\textit{c}): as \textit{a)} but for simulations with an additional $0.2$~mag intrinsic luminosity step at a SN progenitor age of $0.75$~Gyr. \textit{d)}: as \textit{c)} but with SNe split by their host galaxy colour.} 
    
    \label{fig:HR_c_sameRv_1D}
\end{figure*}

\begin{table*}
	\centering
	\caption{Dust population parameters for the models presented in this work when compared to DES5YR data split at host stellar mass $\log({M_*/\mathrm{M}_{\odot}})=10$ (left hand panels of Figs. \ref{fig:HR_c_1D_BS21} \& \ref{fig:HR_c_1D_age}.) $Y$ represents the host galaxy or SN property used to determine $R_V$, while $X$ is the property on which an intrinsic luminosity step (of magnitude $\gamma_X$; Eq. \ref{eq:Gamma}) is placed.}
	\label{tab:res_split_mass}
	\begin{tabular}{lcccccccccccccc} 
		\hline
		Model Name &BiasCor& $\overline{R_{V,1}}$  & $\sigma_{R_V, 1}$  &  $\overline{R_{V,2}}$  & $\sigma_{R_V, 2}$  & $Y$ & $Y_{\mathrm{split}}$ & $Y_1$  & $Y_2$ &$X$ & $X_{\mathrm{split}}$& $\gamma_{X}$ & $\chi^2_{\nu,\mu_{\mathrm{res}}}(M_*)$ \\
		\hline
		BS21 & 1D & 1.75 & 1.0 & 3.0 & 1.0 & $M_*$ & $10^{10}~ \mathrm{M}_{\odot}$ & - & -& $M_*$  &  $10^{10}~ \mathrm{M}_{\odot}$& 0.0  &1.63 \\
		BS21 & 1D & 1.75 & 1.0 & 3.0 & 1.0 &  $M_*$ &   $10^{10}~ \mathrm{M}_{\odot}$ & - & -&$\tau_{\mathrm{A}}$&0.75~Gyr & 0.0  &2.03 \\
	    Age $R_V$ & 1D & 1.5 & 1.0 & 3.0 & 1.0 & $\tau_{\mathrm{G}}$ &  3 ~Gyr & - & - & $\tau_{\mathrm{A}}$&0.75~Gyr &0.0 &1.94\\
        Age $R_V$ & 1D & 1.5 & 1.0 & 2.75 & 1.0 &  $\tau_{\mathrm{G}}$ &  3~Gyr & - & - & $M_*$& $10^{10}~ \mathrm{M}_{\odot}$& 0.0 &1.86\\
        Age $R_V$ Linear & 1D &1.5 & 1.0 & 2.5 & 1.0 & $\tau_{\mathrm{A}}$ & - & 0.1~Gyr & 10~Gyr &$\tau_{\mathrm{A}}$&0.75~Gyr &0.15 &2.27\\	
	    
		\hline
		
	\end{tabular}
\end{table*}

\begin{table*}
	\centering
	\caption{Dust population parameters for the  models presented in this work when compared to DES5YR data split at host $U-R=1$ (right hand panels of Figs. \ref{fig:HR_c_1D_BS21} \& \ref{fig:HR_c_1D_age}.)}
	\label{tab:res_split_UR}
	\begin{tabular}{lcccccccccccccc} 
		\hline
		Model Name &BiasCor& $\overline{R_{V,1}}$  & $\sigma_{R_V, 1}$  &  $\overline{R_{V,2}}$  & $\sigma_{R_V, 2}$  & $Y$ & $Y_{\mathrm{split}}$ & $Y_1$  & $Y_2$ &$X$ & $X_{\mathrm{split}}$& $\gamma_{X}$ & $\chi^2_{\nu,\mu_{\mathrm{res}}}(U-R)$ \\
		\hline
		BS21 & 1D & 1.5 & 1.0 & 3.0 & 1.0 & $M_*$ & $10^{10}~ \mathrm{M}_{\odot}$ & - & -&$M_*$ & $10^{10}~ \mathrm{M}_{\odot}$& 0.10  & 1.70\\
		BS21 & 1D & 1.5 & 1.0 & 3.0 & 1.0 &  $M_*$ &  $10^{10}~ \mathrm{M}_{\odot}$ & - & -&$\tau_{\mathrm{A}}$ &0.75~Gyr & 0.10  &1.59 \\
	    Age $R_V$& 1D & 1.75 & 1.0 & 2.5 & 1.0 & $\tau_{\mathrm{G}}$ & 3~Gyr & - & - &$\tau_{\mathrm{A}}$ &0.75~Gyr  & 0.15 &1.70\\
		Age $R_V$& 1D & 1.5 & 1.0 & 2.5 & 1.0 & $\tau_{\mathrm{G}}$ & 3~Gyr & - & - &$M_*$ &$10^{10}~ \mathrm{M}_{\odot}$  & 0.1 &1.65\\
	    Age $R_V$ Linear & 1D &1.5 & 1.0 & 2.5 & 1.0 & $\tau_{\mathrm{A}}$ & - & 0.1~Gyr & 10~Gyr &$\tau_{\mathrm{A}}$ &0.75~Gyr  & 0.20 &1.99\\

		\hline
		
	\end{tabular}
\end{table*}

\begin{figure*}

	\includegraphics[width=\textwidth]{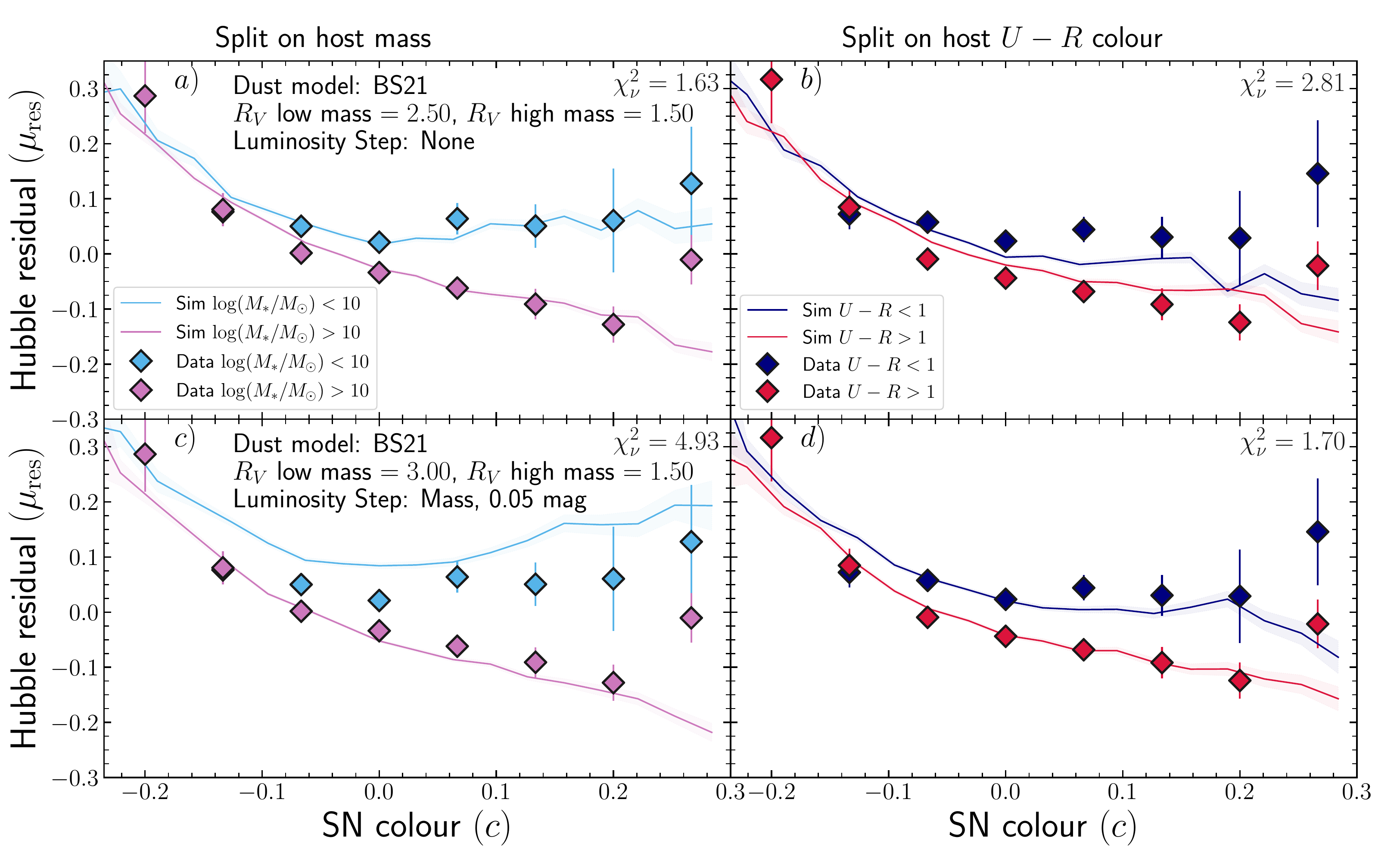}
    \caption{SN Ia Hubble residual versus SN colour, as in Fig. \ref{fig:HR_c_sameRv_1D}, but for models with $R_V$ that varies with host galaxy stellar mass. \textit{a):} Simulations from the BS21 model with $\overline{R_V}=2.5$ in low mass host galaxies and $\overline{R_V}=1.5$ in high mass host galaxies. Hubble residuals for data and simulations have been measured with BBC-1D; \textit{b):} as \textit{a}, but for the data and simulations split host galaxy $U-R = 1$. The model that describes the mass-split data well does not reproduce the full difference between Hubble residuals split on $U-R$.
    \textit{c) \& d):} as \textit{a} \& \textit{b} but for the model parameters that best describe the $U-R$ data (right hand panel): $\overline{R_V}=3.0$ in low mass host galaxies and $\overline{R_V}=1.5$ in high mass host galaxies, and an intrinsic luminosity step at $\log(M/\mathrm{M}_{\odot})=10$ of size $0.1$~mag.}
    \label{fig:HR_c_1D_BS21}
\end{figure*}

\begin{figure*}

	\includegraphics[width=\textwidth]{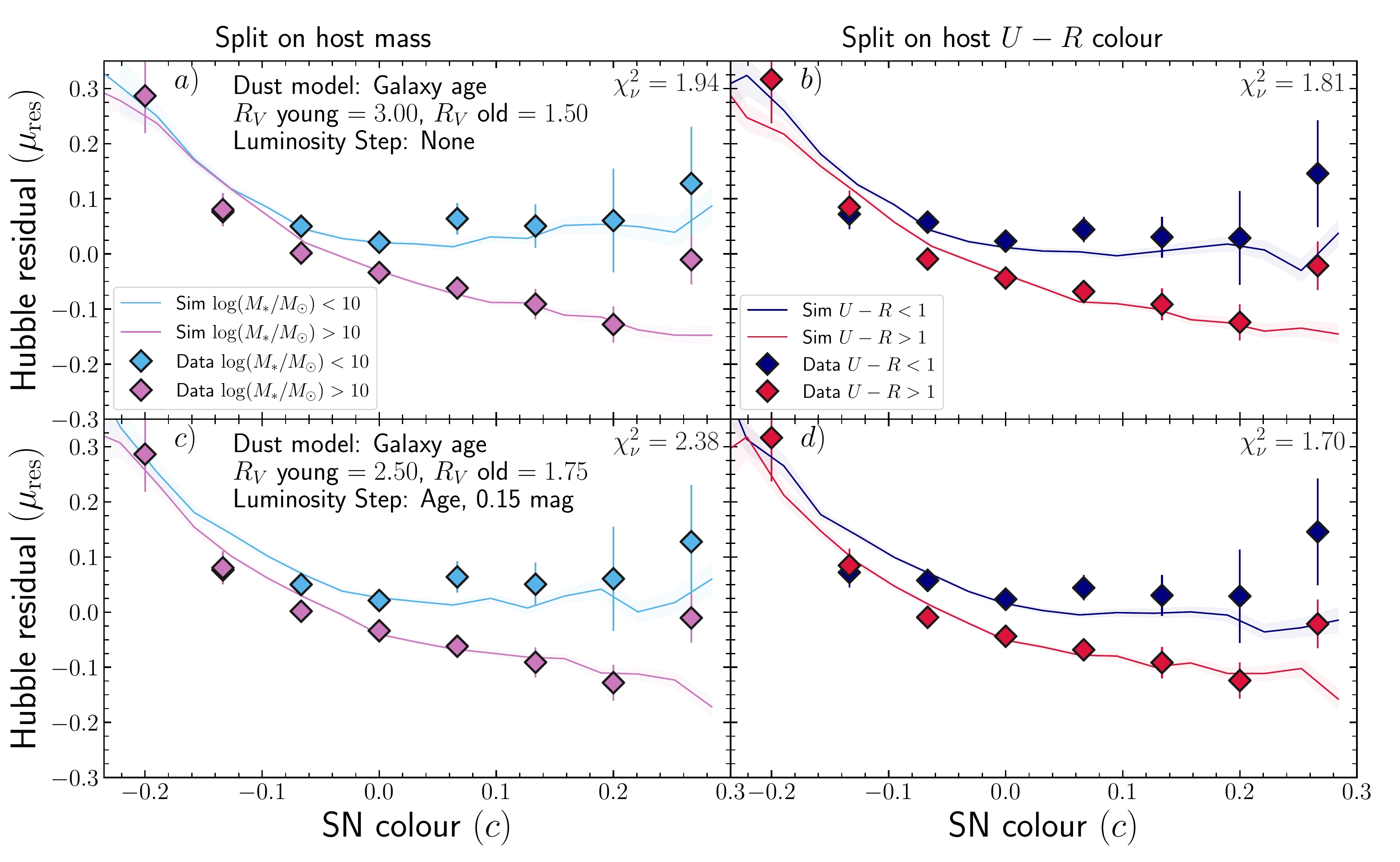}
    \caption{SN Ia Hubble residual versus SN colour, as in Fig. \ref{fig:HR_c_1D_BS21} but for the model with $R_V$ changing with galaxy age. 
    \textit{a)}: The model that best fits the SN data when split by stellar mass includes $\overline{R_V}=3.0$ in young hosts, $\overline{R_V}=1.5$ in old hosts and no intrinsic luminosity step.
    \textit{b)}: as \textit{a}, but showing SNe split at host galaxy $U-R=1$. 
	\textit{c) \& d)}: as for \textit{a \& b} but for the best fitting model to the data split by host $U-R$ colour: $\overline{R_V}=2.5$ in young hosts, $\overline{R_V}=1.75$ in old hosts, and a $0.15$~mag intrinsic luminosity step at a progenitor age of $0.75$~Gyr.}
     \label{fig:HR_c_1D_age}
\end{figure*}

\section{Results}
\label{sec:results}

The principle aim of this work is to explain the observed trends between Hubble residual and SN $c$, with the sample divided by both host galaxy $M_*$ and host galaxy $U-R$ colour. In this section, we present the results of simulating a sample of SN Ia host galaxies (Section \ref{sec:sims_hosts}), light curve properties (Section \ref{subsec:sim_lc}), and fitting their Hubble residuals (Section \ref{subsubsec:sim_HR}) using different assumptions of the generative models for $x_1$ and $R_V$ and various forms of intrinsic luminosity step. 

For each of our model variations, all of the parameters defining the population distributions are held fixed apart from three parameters of interest, for which we simulate a course grid. These parameters are the two extinction law $R_V$ parameters and the luminosity step size. We vary $\overline{R_{V,1}}$ between $[1.5-2.75]$ in steps of 0.25, $\overline{R_{V,2}}$ in the range $[2.5-3.75]$ with steps of 0.25, and $\gamma_{X}$ between $[0-0.2]$ with steps of 0.05. Future analyses will employ inference techniques to precisely constrain the best fit parameters of these models for their use in cosmological analyses, as has been performed for the BS21 model in \citet{Popovic2021a}. The relationship between simulated SN $R_V$ as a function of host galaxy stellar mass and $U-R$ colour is shown in Fig. \ref{fig:Rv_v_gals}, for a representative simulation with $\overline{R_{V,1}}=1.75$, $\overline{R_{V,2}}=3.0$. It is clear that an $R_V$ step in age is better recovered by $U-R$ colour than stellar mass, and that stellar mass steps are not particularly well recovered by splitting colour at $U-R =1$, highlighting the need for both diagnostics.

Each model is compared to the Hubble residuals measured in the DES5YR data of \citetalias{Kelsey2022} split by  $M_*$ and $U-R$. A summary of the results of our simulations and comparison to the data is presented in Tables \ref{tab:res_split_mass} \& \ref{tab:res_split_UR}.

 Since SNe are simulated stochastically by sampling from a number of interlinked distributions and a numerical host galaxy library, an analytical evaluation of the relationship between Hubble residual and $c$ in the model is not possible. We thus compare the simulated and observed trends in a similar way to the one-dimensional parameter distributions (Section \ref{subsec:res_valid_lcs}).
 
 For each bin $i$ in a host galaxy property $X$, we group simulated SNe in the same bins $c_j$ as used for the data. The reduced chi-squared is then
\begin{equation}
    \chi^2_{\nu,~\mu_{\mathrm{res}}} = \frac{1}{N_{\mathrm{bins}}} \sum_i^X \sum_j^{N_{\mathrm{bins}}} \frac{\left(\overline{\mu_{\mathrm{res,data},i,j}} - \overline{\mu_{\mathrm{res,sim},i,j}}\right)^2}{e^2_{i,j}}\,,
\label{eq:chi2_HR_c}
\end{equation}
where $\overline{\mu_{\mathrm{res,data},i,j}}$ and $\overline{\mu_{\mathrm{res,sim},i,j}}$ are the weighted mean Hubble residuals in each host bin $i$ and colour bin $j$ for data and simulations, respectively, and $e_{i,j}$ is the standard error on the means of the data.

Each of the following sections present the results for one model of how $R_V$ varies with host or SN properties. For each model we present the data and simulations (Figs. \ref{fig:HR_c_sameRv_1D}-\ref{fig:HR_c_1D_age}) split by their stellar mass (left-hand panels) and $U-R$ (right-hand panels). The values from our grid search that result in the smallest $\chi^2_{\nu,~\mu_{\mathrm{res}}}$ for each model when split by stellar mass and $U-R$ are presented in Tables \ref{tab:res_split_mass} and \ref{tab:res_split_UR} respectively. Unless otherwise stated, parameters reported below are determined using a 1D (BBC-1D) bias correction. Corresponding figures and tables for BBC-BS20 can be found in Appendix \ref{sec:app_4D}. We stress that the $R_V$ values here are indicative only, since the generative models for the simulated data and bias corrections are not the same. A thorough fitting via the method of \citet{Popovic2021a} is left for future work.

\subsection{Fixed $R_V$, no luminosity step}
\label{subsec:res_fixRv}

We begin by simulating SNe from a baseline model with no luminosity step and $\overline{R_V}$ fixed at 2.5 across all host galaxies.  The resulting Hubble residuals shown in Fig. \ref{fig:HR_c_sameRv_1D} (panels \textit{a, b}) display a trend with SN colour $c$, as expected due to the effects of parameter migration \citep{Scolnic2016}, but they do not reproduce the differences between low and high stellar mass, or between blue and red host galaxies. To test whether a simple luminosity step as a function of SN age can explain the diverging Hubble residuals, we add a luminosity age step $\gamma_{\tau_A} = 0.2$~mag to the simulation (Fig. \ref{fig:HR_c_sameRv_1D}, panels \textit{c, d}). Again, the diverging Hubble residuals are not reproduced. 
Notably the simulated $0.2$~mag luminosity step is not recovered in the Hubble residuals, with a step of only around $0.05~$ mag appearing. The reasons behind this are discussed in Section \ref{subsec:disc_steps}.

\subsection{$R_V$ split on stellar mass (BS21)}
\label{subsec:res_BS21}

Motivated by the clear signal in the data of diverging Hubble residuals as a function of SN $c$, we continue by implementing the model of \citetalias{Brout2020}. Our \citetalias{Brout2020} simulation is shown in Fig. \ref{fig:HR_c_1D_BS21} (upper panels), and qualitatively replicates the trend in the data when split by stellar mass. Simulations with $\overline{R_V}=3.0$ in low mass galaxies and $1.75$ in high mass hosts match the data well ($\chi^2_{\nu}=1.63$), values that are consistent with those found by \citetalias{Brout2020} and \citet{Popovic2021a}. 

However, this model that describes the mass-split data adequately performs less well when the data are split by host $U-R$ colour ($\chi^2_{\nu}=2.81$), and neither the difference between Hubble residuals in blue nor red SNe are matched as well. The best match $(\chi^2_{\nu}=1.70)$, shown in the lower panels of Fig. \ref{fig:HR_c_1D_BS21}, has a steeper extinction law in high mass galaxies with $\overline{R_V} = 1.5$, as well as the addition of an intrinsic $0.1$~mag luminosity step on stellar mass. However, the addition of this step has the effect of drastically reducing the quality of the match when the data are split by stellar mass $(\chi^2_{\nu}=4.93)$.

\subsection{$R_V$ split on galaxy age}
\label{subsec:res_age_BS21}

The \citetalias{Brout2020} model is able to reproduce the observed trends between Hubble residual and SN $c$ when split by stellar mass but the model requires different parameters and an intrinsic luminosity step when matching the data split by host $U-R$ colour. Motivated by a hypothesis that dust parameters could be driven by stellar age rather than galaxy mass, we implement the \lq Age $R_V$\rq\ model, which is identical to the \citetalias{Brout2020} model but with $R_V$ described by a step function at a mean stellar age of 3~Gyr rather than $\log(M/\mathrm{M}_{\odot})=10$. This model also incorporates the age-based \citet{Nicolas2020} model of $x_1$.

As with the mass-based model, the better matching simulations are those with differing $\overline{R_V}$ in young and old galaxies, with larger differences in $\overline{R_V}$ required compared to the \citetalias{Brout2020} model. When splitting Hubble residuals on stellar mass (Fig. \ref{fig:HR_c_1D_age} upper panels, Table \ref{tab:res_split_mass}), the favoured models  $(\chi^2_{\nu}=1.94)$ include a similar difference in $R_V$ between old (1.5) and young (2.75) hosts and no intrinsic luminosity step is required.
When splitting the data on $U-R$ this model provides a significantly better match than the mass-$R_V$-step model of \citetalias{Brout2020} (when implemented with no further luminosity step, i.e. Fig. \ref{fig:HR_c_1D_age}b has $\chi^2_{\nu}=1.70$ compared to Fig. \ref{fig:HR_c_1D_BS21}b with $\chi^2_{\nu}=2.44$). The best match to the $U-R$ split data has a smaller difference between $R_V$ values (1.5 and 2.5) plus additional 0.1~mag intrinsic luminosity step on stellar mass, or an even smaller $R_V$ difference (1.75 and 2.5) and a 0.15~mag step on stellar age (Fig. \ref{fig:HR_c_1D_age}, lower panels). As described in the results of the previous models, a much smaller step is evident in the Hubble residuals than the age step input into the SNe. We address this effect in Section \ref{subsec:disc_steps}.

\subsection{$R_V$ linear with SN age}
\label{subsec:res_age_linear}

Modelling $R_V$ as a linear function of SN progenitor age is not as successful as the previous step-based models, with consistently higher $\chi^2_{\nu}$ (Tables~\ref{tab:res_split_mass} \& \ref{tab:res_split_UR}). This $R_V$ model is the only one that also requires an age-luminosity step to explain both the mass-split and $U-R$ split data, with $\gamma_X \geq 0.15$~mag.

\begin{figure*}

	\includegraphics[width=\textwidth]{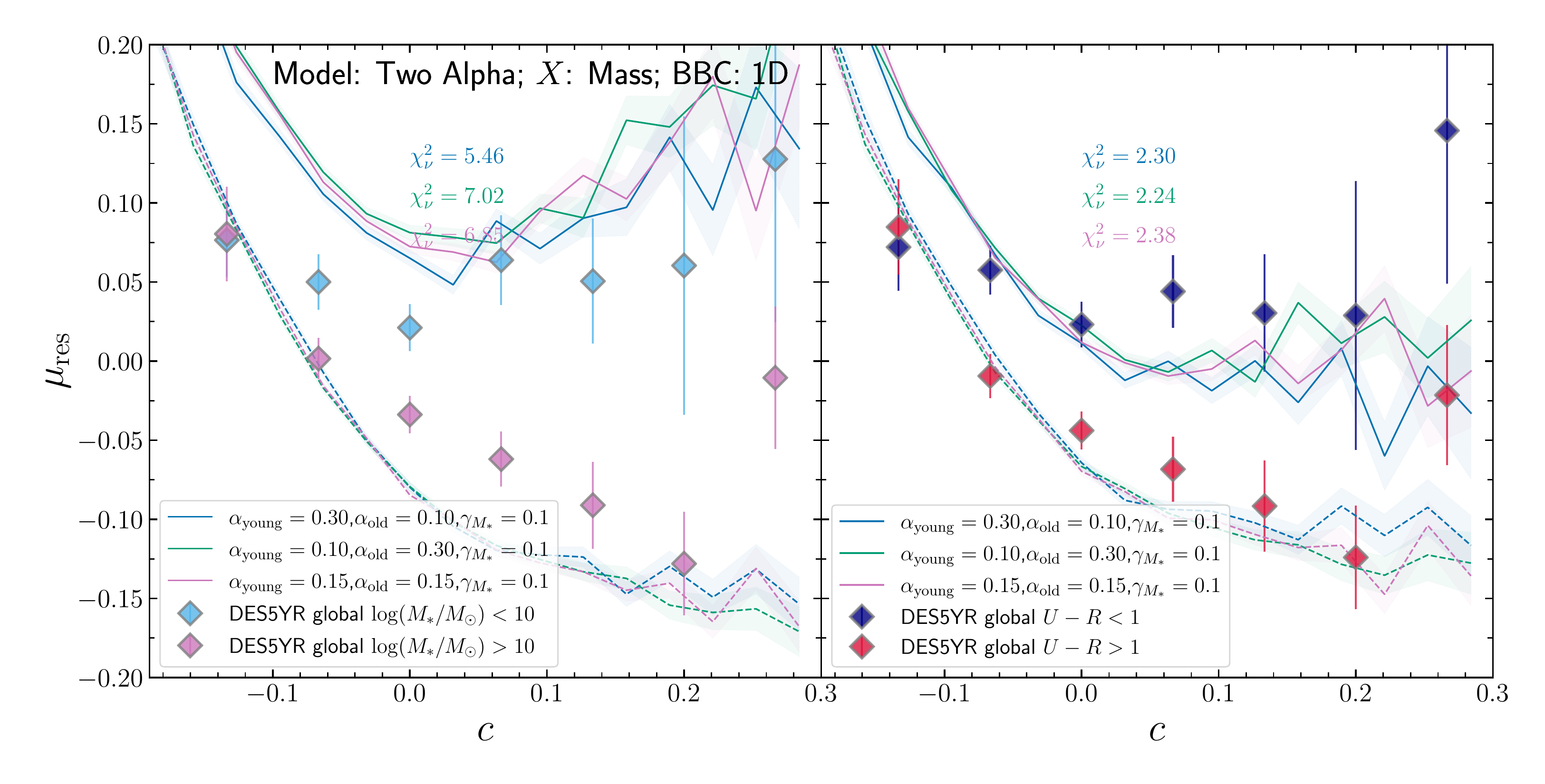}
    \caption{\textit{Left}: As Fig. \ref{fig:HR_c_1D_age} with fixed $R_V$ ($R_{V,1}=2.0$; $R_{V,2}=3.75$, $Y= \tau_\mathrm{G}$), an intrinsic mass step $\gamma_{M_*}=0.1$~mag, and varying $\alpha_{\mathrm{sim}}$ for young and old populations of SNe. There is no significant difference in the trend of HRs with $c$, nor the size of the measured mass step, despite large variations in $\alpha_{\mathrm{sim}}$; \textit{Right}: as left, but for host galaxy $U-R$ colour. As with mass, there is no significant difference in the HRs despite large variations in $\alpha_{\mathrm{sim}}$ and the recovered $U-R$ step is not affected. }
    \label{fig:two_alphas}
\end{figure*}

\section{Discussion}
\label{sec:discussion}

The results of our comparisons of simulations to data support the notion that the SN colour--luminosity relation in SNe Ia is not linear with SN colour, consistent with the model of \citetalias{Brout2020} where the dust extinction law along the line-of-sight to SNe Ia correlates with the global properties of host galaxies. The reason for, and best tracer of, the difference in $R_V$ is less clear and is discussed in the following sections. As with the previous section, the discussion here focuses on BBC-1D bias corrections, but the results are consistent when using BBC-BS20 bias corrections.

\subsection{$R_V$ in young, low mass galaxies}
\label{subsec:discussion_highRv}

We find no significant difference in the fit quality when the $R_V$ changes with stellar mass or stellar age, the two of which are themselves strongly correlated. When implementing the \citetalias{Brout2020} model we find $\overline{R_{V,2}}$ is broadly consistent with the results from \citetalias{Brout2020} and \citet{Popovic2021a} when splitting on stellar mass. When $R_V$ changes with stellar age the simulations are more consistent with the data across both stellar mass and $U-R$, but $2.5 \leq \overline{R_{V,2}} \leq3.0$ are generally smaller than the 3.0 in \citet{Popovic2021a}. These values for $R_V$ consistent with those typically measured in SN Ia light curves and spectra for similar hosts, with \citet{Cikota2016} finding $R_V$ values of $2.71\pm1.58$ in spiral galaxies. 


\subsection{$R_V$ in old, high mass galaxies}

In all of the models trialled the $R_V$ value in high mass/old galaxies ($\overline{R_{V,1}}$) is required to be significantly lower than that for low mass/young galaxies in order to replicate the trends of Hubble residual against SN colour. Generally, lower values are preferred when the data are split on $U-R$. Values in the range $1.5 \leq \overline{R_{V,1}} \leq 1.75$ indicate the dust in older, more massive SN hosts is composed mainly of small grains leading to a steep extinction curve. How this observation relates to studies of the general galaxy population is complicated by different relationships seen between dust and different galaxy descriptors. While \citet{Reddy2018} and \citet{Salim2018} show that $R_V$ increases with stellar mass for star-forming galaxies, \citet{Salim2018} also show that $R_V$ is lower in passive (and starburst) galaxies than those on the star forming main sequence, and that the $R_V$ of star-forming galaxies is more closely related to its overall extinction $A_V$ than its stellar mass. However, the quiescent galaxies in \citet{Salim2018} still have an average $R_V$ of 2.61, far greater than that found here and indicating that the SNe do not trace the average dust composition in the galaxy. We note however that the SN $R_V$ measurements are somewhat dependent on the assumed SALT2 colour law, rendering direct comparisons less informative.

\subsection{The effect of intrinsic luminosity steps}
\label{subsec:disc_steps}
While models of an $R_V$ step in stellar mass or galaxy age provide reasonable matches to the Hubble residuals split by either stellar mass or host $U-R$ colour, they do not fully match both host properties simultaneously -- different $R_V$ steps are required to match mass and to match $U-R$.
Furthermore, \citetalias{Kelsey2022} show that there are residual steps in the data in one parameter (e.g. stellar mass) if the effect of another (e.g. $U-R$) is subtracted. We find that in general the step seen in our simulated HRs is smaller in $U-R$ than in $M_*$, but that step is rarely fully removed using $R_V$ only and that adding an intrinsic luminosity step is required. However, such a step introduces a number of its own issues.

Firstly we consider adding a step based on the SN age. We find that adding a 0.2~mag step has only a marginal effect on the HRs -- instead, the best fit value $\alpha$ is much smaller than the simulated value $\alpha_{\mathrm{sim}}$. This effect is caused in this model by the dependence on age of both the intrinsic step and $x_1$, which act on the SN luminosity in the opposite direction to each other. Such an affect has been observed before \citep[e.g.][]{Rose2021}, who suggest that steps $\gamma_X$ should be fit simultaneously with the other nuisance parameters. However, with the step size evolving with SN colour, this is not trivial. 

An intrinsic mass step is affected much less absorbed by $\alpha$, with a $0.1~$mag step able to explain the $U-R$ data for most $R_V$ model and bias correction combinations (e.g. Fig. \ref{fig:HR_c_1D_BS21}). However, this means that the discrepancy between models with an intrinsic mass step and the data split by mass is much larger than when the intrinsic step is on age and no particularly good combinations are found.

\subsubsection{Varying the width-luminosity coefficient $\alpha$}
\label{subsubsec:alpha}

Neither an intrinsic luminosity step based on SN age or host mass are able to simultaneously model the data split by stellar mass and by $U-R$. One logical explanation for this is that the relationships between SN age, $x_1$, and $\alpha$ are linked with different strengths to the host parameters $M_*$ and $U-R$. The $U-R$ colour traces age more directly than $M_*$, and with the bimodal \citetalias{Nicolas2020} model $x_1$ is strongly linked to SN age, meaning intrinsic steps in luminosity are washed out by being absorbed into the alpha parameter, and this occurs more strongly in the age-like $U-R$ than $M_*$ which is less related to SN age. To counter this effect, we hypothesise that the two $x_1$ populations could follow different intrinsic values of $\alpha$ while simultaneously having different mean absolute magnitudes.

To test the two-$\alpha$ hypothesis we run a further set of simulations. We fix $R_V$ to those from the best Age-$R_V$ model as measured with BBC-1D and as compared to data split by mass: ($R_{V,1}=1.5$, $R_{V,2}=3.0$). We simulate over a new grid, varying $\alpha_{\mathrm{sim}}$ from $0.05$ to $0.30$ for the young and old populations, as well as the intrinsic luminosity step $\gamma_X$. The combinations of $\alpha_{\mathrm{sim}}$ and $\gamma_X$ that we investigate are constrained by the fitted $\alpha$. For $X$, we trial both mass and age steps between zero and $0.25$~mag. 

The results of varying $\alpha_{\mathrm{sim}}$ are summarised in Fig. \ref{fig:two_alphas}, for which the models included a 0.1~mag step on $M_*$. Varying $\alpha_{\mathrm{sim}}$ for the young and old SNe makes no difference to the inferred step in either $M_*$ or $U-R$ and does not solve the discrepancy between the size of recovered steps -- whatever the values of $\alpha_{\mathrm{sim}}$, the resulting HRs are split much more strongly by $M_*$ than by $U-R$. The same results were found when including a 0.2~mag luminosity step on SN age. These simulations thus indicate a universal value for $\alpha$ even if the SNe are divided into two populations of $x_1$. Such a lack of bimodality for $\alpha$ is consistent with previous results, e.g. \citet{Sullivan2011}.

\subsubsection{Reconciling mass and $U-R$ steps}

With multiple $\alpha$ values improbable, we investigate whether the discrepancy is caused by limitations in our model. For example, the smooth SFHs lead to a single Gaussian distribution of $U-R$ while the data show bimodality, while the simulation also slightly underestimates the number of hosts in the stellar mass range $9\leq\log(M/\mathrm{M}_{\odot})$. Although Fig. \ref{fig:mass_UR} shows that the models do reproduce the $M_*$--$U-R$ relation in general, there may be subtle second order effects introducing inconsistencies for individual simulated SNe. Such effects include the treatment of metallicity which is fixed at solar in the simulation but evolves over both the mass and redshift range of the data \citep[e.g.][]{Tremonti2004,Zahid2014}, the lack of bursts of star formation in the simulation, the choice of BC03 templates for the simulation, and incorrect modelling of survey selection effects. These effects will be investigated in subsequent work. 

A second possibility is that the inconsistencies between the model and data are caused by physical effects that have not been fully incorporated or accurately implemented in the model. The relationships between the $R_V$ along a SN sight line and its host galaxy properties are evidently more complicated that a simple step function of galaxy age or stellar mass -- when dealing with integrated galaxy attenuation, $R_V$ is inversely correlated with $A_V$, which we have not included in the model for SN $R_V$. Meanwhile, $A_V$ correlates strongly with stellar mass \citep{Zahid2013}, meaning the $R_V$ of galaxies also increases with stellar mass \citep{Salim2018}. On the contrary, the results of \citetalias{Brout2020}, \citet{Popovic2021a}, and this work, indicate that $R_V$ on SN sight lines decreases with stellar mass, suggesting that SN sight lines are systematically different than those of the integrated galaxy light. Meanwhile, \citet{Salim2018} also show that galaxies with old stellar populations have lower $R_V$ than those with young populations, which is in line with the age -- $R_V$ steps found here. Because our the models track individual stellar populations, it is possible to use our method to compare the predictions to measurements from integral field spectroscopy \citep[e.g.][]{Galbany2014,Galbany2016,Galbany2018} to measure dust properties in the very local environments of SNe, which we defer to future work.

Our model has no implementation of a relation between galaxy mass and metallicity. Metallicity is strongly correlated with stellar mass and more weakly with SFR \citep[e.g.][]{Yates2012}, is tied to the SFH \citep{Bellstedt2020} as well as affecting the strengths of nebular emission lines \citep[e.g.][]{Kewley2008}, themselves affecting the $U-R$ colour. If there is an intrinsic correlation between metallicity and SN luminosity \citep[e.g.][]{Hoflich1998,Kasen2009,Moreno-Raya2016} it may affect the observed mass and age steps in different amounts, but be hidden amongst the more dominant $R_V$ effects.

\section{Conclusions and future work}
\label{sec:conclusion}

This work presents a host-galaxy oriented framework for simulating populations of SNe Ia in a cosmological context. We trace stellar population ages throughout the build up of stellar mass of galaxies, and use this information combined with the SN Ia DTD from \citetalias{Wiseman2021} in order to associate SNe to hosts at realistic rates. The resulting host galaxy library is made available to the community for use in future cosmological SN Ia simulations\footnote{\url{https://github.com/wisemanp/des_sn_hosts/tree/main/simulations/data}}. SNe are then simulated according to their host galaxy properties as highlighted below:

\begin{itemize}
    \item light curve width $x_1$ is drawn from a two-population age-based model based on \citetalias{Nicolas2020}, which accurate reproduces the $x_1$ distribution. The $x_1$ vs stellar mass and $x_1$ vs $U-R$ relations are also relatively well modelled, although the strength of the relation is stronger in the data than we recover in the model. 
    \item we that 68\% of SNe in old environments belong to the low-stretch mode. Further work is necessary in order to determine the cause of these two modes, whether they are related to the progenitor scenario, white dwarf composition, or explosion mechanism, and why the transition occurs around 0.75~Gyr;
    
    \item SN colour $c$ is well described by a combination of an intrinsic Gaussian and a supplementary exponential distribution attributed to dust reddening. The DES5YR data is best modelled by very similar values to those for the Pantheon+ dataset \citep{Popovic2021a}.
    
\end{itemize}
By running simulations and data through the same BBC framework we obtain distance estimates and compare the evolution of the Hubble residuals with $c$. Our results support the findings of \citetalias{Brout2020} and \citet{Popovic2021a} that the extinction law slope $R_V$ changes depending on host galaxy properties. Additionally we find that:

\begin{itemize}
    \item if galaxy age is the driver of $R_V$ change, then the evolution is different to if it is driven by stellar mass. However the change (with mass) is opposite to that observed in the galaxy population, while the change with age is consistent with the difference between star forming and passive galaxies;
    
    \item when Hubble residuals are split based on host stellar mass, the $R_V$ models can fully explain the trends between Hubble residual and $c$;
    
    \item when the data are split by host $U-R$, the addition of an intrinsic luminosity step to the model slightly improves the fit to the data. 
    
    \item there is no preference for the intrinsic luminosity step to be based on stellar mass or SN age, but neither are able to simultaneously reproduce the data when split by host stellar mass and by host $U-R$. 
    
    \item varying the width-luminosity coefficient $\alpha$ between populations does not solve the discrepancy. 
    
The reason for the discrepancy is therefore either a shortcoming of the simulation or an unmodelled physical effect such as metallicity.
\end{itemize}

Future work will build upon the host galaxy models developed here. In order to test for effects such as metallicity, the galaxy evolution model needs to be treated in a more complex way by introducing metallicity evolution and bursts of star formation. Such models have been used in the modelling of SN Ia rates and DTDs \citep[e.g.][]{Gandhi2022} and can be extended to cosmological implementation using the methods outlined in this paper.

\section*{Software}
All software used in this publication are publicly available. The SN and galaxy evolution code can be found at https://github.com/wisemanp/des\_sn\_hosts. Additionally we made extensive use of numpy \citep{Harris2020}, Astropy \citep{AstropyCollaboration2018}, matplotlib \citep{Hunter2007}, SciPy \citep{Virtanen2020}, and pandas \citep{Mckinney2010}.

\section*{Acknowledgements}
We thank the referee Mickael Rigault for comments which significantly improved this paper.
P.W. acknowledges support from the Science and Technology Facilities Council (STFC) grant ST/R000506/1. M.S acknowledges support from EU/FP7-ERC grant 615929. L.K. thanks the UKRI Future Leaders Fellowship for support through the grant MR/T01881X/1. L.G. acknowledges financial support from the Spanish Ministerio de Ciencia e Innovaci\'on (MCIN), the Agencia Estatal de Investigaci\'on (AEI) 10.13039/501100011033, and the European Social Fund (ESF) "Investing in your future" under the 2019 Ram\'on y Cajal program RYC2019-027683-I and the PID2020-115253GA-I00 HOSTFLOWS project, from Centro Superior de Investigaciones Cient\'ificas (CSIC) under the PIE project 20215AT016, and the program Unidad de Excelencia Mar\'ia de Maeztu CEX2020-001058-M. This project has received funding from the European Research Council (ERC) under the European Union's Horizon 2020 research and innovation program (grant agreement number 759194 - USNAC).

Funding for the DES Projects has been provided by the U.S. Department of Energy, the U.S. National Science Foundation, the Ministry of Science and Education of Spain, 
the Science and Technology Facilities Council of the United Kingdom, the Higher Education Funding Council for England, the National Center for Supercomputing 
Applications at the University of Illinois at Urbana-Champaign, the Kavli Institute of Cosmological Physics at the University of Chicago, 
the Center for Cosmology and Astro-Particle Physics at the Ohio State University,
the Mitchell Institute for Fundamental Physics and Astronomy at Texas A\&M University, Financiadora de Estudos e Projetos, 
Funda{\c c}{\~a}o Carlos Chagas Filho de Amparo {\`a} Pesquisa do Estado do Rio de Janeiro, Conselho Nacional de Desenvolvimento Cient{\'i}fico e Tecnol{\'o}gico and 
the Minist{\'e}rio da Ci{\^e}ncia, Tecnologia e Inova{\c c}{\~a}o, the Deutsche Forschungsgemeinschaft and the Collaborating Institutions in the Dark Energy Survey. 

The Collaborating Institutions are Argonne National Laboratory, the University of California at Santa Cruz, the University of Cambridge, Centro de Investigaciones Energ{\'e}ticas, 
Medioambientales y Tecnol{\'o}gicas-Madrid, the University of Chicago, University College London, the DES-Brazil Consortium, the University of Edinburgh, 
the Eidgen{\"o}ssische Technische Hochschule (ETH) Z{\"u}rich, 
Fermi National Accelerator Laboratory, the University of Illinois at Urbana-Champaign, the Institut de Ci{\`e}ncies de l'Espai (IEEC/CSIC), 
the Institut de F{\'i}sica d'Altes Energies, Lawrence Berkeley National Laboratory, the Ludwig-Maximilians Universit{\"a}t M{\"u}nchen and the associated Excellence Cluster Universe, 
the University of Michigan, NSF's NOIRLab, the University of Nottingham, The Ohio State University, the University of Pennsylvania, the University of Portsmouth, 
SLAC National Accelerator Laboratory, Stanford University, the University of Sussex, Texas A\&M University, and the OzDES Membership Consortium.

Based in part on observations at Cerro Tololo Inter-American Observatory at NSF's NOIRLab (NOIRLab Prop. ID 2012B-0001; PI: J. Frieman), which is managed by the Association of Universities for Research in Astronomy (AURA) under a cooperative agreement with the National Science Foundation.

The DES data management system is supported by the National Science Foundation under Grant Numbers AST-1138766 and AST-1536171.
The DES participants from Spanish institutions are partially supported by MICINN under grants ESP2017-89838, PGC2018-094773, PGC2018-102021, SEV-2016-0588, SEV-2016-0597, and MDM-2015-0509, some of which include ERDF funds from the European Union. IFAE is partially funded by the CERCA program of the Generalitat de Catalunya.
Research leading to these results has received funding from the European Research
Council under the European Union's Seventh Framework Program (FP7/2007-2013) including ERC grant agreements 240672, 291329, and 306478.
We  acknowledge support from the Brazilian Instituto Nacional de Ci\^encia
e Tecnologia (INCT) do e-Universo (CNPq grant 465376/2014-2).

This manuscript has been authored by Fermi Research Alliance, LLC under Contract No. DE-AC02-07CH11359 with the U.S. Department of Energy, Office of Science, Office of High Energy Physics.

\section*{Data Availability}
The light curves for the DES-SN photometric SN Ia catalogue and associated host galaxy data will be made available as part of the DES5YR SN cosmology analysis at https://des.ncsa.illinois.edu/releases/sn.



\bibliographystyle{mnras}
\bibliography{PhilMendeley} 

\begin{thebibliography}{}
\makeatletter
\relax
\def\mn@urlcharsother{\let\do\@makeother \do\$\do\&\do\#\do\^\do\_\do\%\do\~}
\def\mn@doi{\begingroup\mn@urlcharsother \@ifnextchar [ {\mn@doi@}
  {\mn@doi@[]}}
\def\mn@doi@[#1]#2{\def\@tempa{#1}\ifx\@tempa\@empty \href
  {http://dx.doi.org/#2} {doi:#2}\else \href {http://dx.doi.org/#2} {#1}\fi
  \endgroup}
\def\mn@eprint#1#2{\mn@eprint@#1:#2::\@nil}
\def\mn@eprint@arXiv#1{\href {http://arxiv.org/abs/#1} {{\tt arXiv:#1}}}
\def\mn@eprint@dblp#1{\href {http://dblp.uni-trier.de/rec/bibtex/#1.xml}
  {dblp:#1}}
\def\mn@eprint@#1:#2:#3:#4\@nil{\def\@tempa {#1}\def\@tempb {#2}\def\@tempc
  {#3}\ifx \@tempc \@empty \let \@tempc \@tempb \let \@tempb \@tempa \fi \ifx
  \@tempb \@empty \def\@tempb {arXiv}\fi \@ifundefined
  {mn@eprint@\@tempb}{\@tempb:\@tempc}{\expandafter \expandafter \csname
  mn@eprint@\@tempb\endcsname \expandafter{\@tempc}}}

\bibitem[\protect\citeauthoryear{{Astropy Collaboration} et~al.,}{{Astropy
  Collaboration} et~al.}{2018}]{AstropyCollaboration2018}
{Astropy Collaboration} et~al., 2018, \mn@doi [The Astronomical Journal]
  {10.3847/1538-3881/aabc4f}, 156, 123

\bibitem[\protect\citeauthoryear{Bellstedt et~al.,}{Bellstedt
  et~al.}{2020}]{Bellstedt2020}
Bellstedt S.,  et~al., 2020, \mn@doi [Monthly Notices of the Royal Astronomical
  Society] {10.1093/mnras/staa2620}, 498, 5581

\bibitem[\protect\citeauthoryear{Bertelli et~al.,}{Bertelli
  et~al.}{1994}]{Bertelli1994}
Bertelli G.,  et~al., 1994, Astronomy and Astrophysics Supplement Series, 106,
  275

\bibitem[\protect\citeauthoryear{Bessell}{Bessell}{1990}]{Bessell1990}
Bessell M.~S.,  1990, \mn@doi [Publications of the Astronomical Society of the
  Pacific] {10.1086/132749}, 102, 1181

\bibitem[\protect\citeauthoryear{Betoule et~al.,}{Betoule
  et~al.}{2014}]{Betoule2014}
Betoule M.,  et~al., 2014, \mn@doi [Astronomy & Astrophysics]
  {10.1051/0004-6361/201423413}, 568, A22

\bibitem[\protect\citeauthoryear{Boquien, Burgarella, Roehlly, Buat, Ciesla,
  Corre, Inoue  \& Salas}{Boquien et~al.}{2019}]{Boquien2019}
Boquien M.,  Burgarella D.,  Roehlly Y.,  Buat V.,  Ciesla L.,  Corre D.,
  Inoue A.~K.,   Salas H.,  2019, \mn@doi [Astronomy & Astrophysics]
  {10.1051/0004-6361/201834156}, 622, A103

\bibitem[\protect\citeauthoryear{Briday et~al.,}{Briday
  et~al.}{2022}]{Briday2021}
Briday M.,  et~al., 2022, \mn@doi [Astronomy & Astrophysics]
  {10.1051/0004-6361/202141160}, 657, A22

\bibitem[\protect\citeauthoryear{Brout \& Scolnic}{Brout \&
  Scolnic}{2021}]{Brout2020}
Brout D.,  Scolnic D.,  2021, \mn@doi [The Astrophysical Journal]
  {10.3847/1538-4357/abd69b}, 909, 26

\bibitem[\protect\citeauthoryear{Brout et~al.,}{Brout
  et~al.}{2019a}]{Brout2019}
Brout D.,  et~al., 2019a, \mn@doi [The Astrophysical Journal]
  {10.3847/1538-4357/ab06c1}, 874, 106

\bibitem[\protect\citeauthoryear{Brout et~al.,}{Brout
  et~al.}{2019b}]{Brout2019a}
Brout D.,  et~al., 2019b, \mn@doi [The Astrophysical Journal]
  {10.3847/1538-4357/ab08a0}, 874, 150

\bibitem[\protect\citeauthoryear{Brout et~al.,}{Brout et~al.}{2022}]{Brout2022}
Brout D.,  et~al., 2022, eprint arxiv:2202.04077

\bibitem[\protect\citeauthoryear{Bruzual \& Charlot}{Bruzual \&
  Charlot}{2003}]{Bruzual2003}
Bruzual G.,  Charlot S.,  2003, \mn@doi [Monthly Notices of the Royal
  Astronomical Society] {10.1046/j.1365-8711.2003.06897.x}, 344, 1000

\bibitem[\protect\citeauthoryear{Cardelli, Clayton  \& Mathis}{Cardelli
  et~al.}{1989}]{Cardelli1989}
Cardelli J.~A.,  Clayton G.~C.,   Mathis J.~S.,  1989, \mn@doi [The
  Astrophysical Journal] {10.1086/167900}, 345, 245

\bibitem[\protect\citeauthoryear{Chabrier}{Chabrier}{2003}]{Chabrier2003}
Chabrier G.,  2003, \mn@doi [Publications of the Astronomical Society of the
  Pacific] {10.1086/376392}, 115, 763

\bibitem[\protect\citeauthoryear{Childress et~al.,}{Childress
  et~al.}{2013}]{Childress2013a}
Childress M.,  et~al., 2013, \mn@doi [Astrophysical Journal]
  {10.1088/0004-637X/770/2/108}, 770, 108

\bibitem[\protect\citeauthoryear{Childress, Wolf  \& Zahid}{Childress
  et~al.}{2014}]{Childress2014}
Childress M.~J.,  Wolf C.,   Zahid H.~J.,  2014, \mn@doi [Monthly Notices of
  the Royal Astronomical Society] {10.1093/mnras/stu1892}, 445, 1898

\bibitem[\protect\citeauthoryear{Cikota, Deustua  \& Marleau}{Cikota
  et~al.}{2016}]{Cikota2016}
Cikota A.,  Deustua S.,   Marleau F.,  2016, \mn@doi [The Astrophysical
  Journal] {10.3847/0004-637x/819/2/152}, 819, 152

\bibitem[\protect\citeauthoryear{{DES Collaboration} et~al.,}{{DES
  Collaboration} et~al.}{2018}]{DESCollaboration2018a}
{DES Collaboration} et~al., 2018, \mn@doi [Physical Review Letters]
  {10.1103/PhysRevLett.122.171301}, 122, 171301

\bibitem[\protect\citeauthoryear{Dixon}{Dixon}{2021}]{Dixon2021}
Dixon S.,  2021, \mn@doi [Publications of the Astronomical Society of the
  Pacific] {10.1088/1538-3873/abef78}, 133, 054501

\bibitem[\protect\citeauthoryear{Dixon et~al.,}{Dixon et~al.}{2022}]{Dixon2022}
Dixon M.,  et~al., 2022, eprint arXiv: 2206.12085

\bibitem[\protect\citeauthoryear{Flaugher et~al.,}{Flaugher
  et~al.}{2015}]{Flaugher2015}
Flaugher B.,  et~al., 2015, \mn@doi [The Astronomical Journal]
  {10.1088/0004-6256/150/5/150}, 150, 150

\bibitem[\protect\citeauthoryear{Galbany et~al.,}{Galbany
  et~al.}{2014}]{Galbany2014}
Galbany L.,  et~al., 2014, \mn@doi [Astronomy & Astrophysics]
  {10.1051/0004-6361/201424717}, 572, 22

\bibitem[\protect\citeauthoryear{Galbany et~al.,}{Galbany
  et~al.}{2016}]{Galbany2016}
Galbany L.,  et~al., 2016, \mn@doi [Monthly Notices of the Royal Astronomical
  Society] {10.1093/MNRAS/STV2620}, 455, 4087

\bibitem[\protect\citeauthoryear{Galbany et~al.,}{Galbany
  et~al.}{2018}]{Galbany2018}
Galbany L.,  et~al., 2018, \mn@doi [The Astrophysical Journal]
  {10.3847/1538-4357/aaaf20}, 855, 107

\bibitem[\protect\citeauthoryear{Gallagher et~al.,}{Gallagher
  et~al.}{2005}]{Gallagher2005}
Gallagher J.~S.,  et~al., 2005, \mn@doi [The Astrophysical Journal]
  {10.1086/491664}, 634, 210

\bibitem[\protect\citeauthoryear{Gandhi, Wetzel, Hopkins, Shappee, Wheeler  \&
  Faucher-Gigu{\`{e}}re}{Gandhi et~al.}{2022}]{Gandhi2022}
Gandhi P.~J.,  Wetzel A.,  Hopkins P.~F.,  Shappee B.~J.,  Wheeler C.,
  Faucher-Gigu{\`{e}}re C.-A.,  2022, eprint arxiv:2202.10477

\bibitem[\protect\citeauthoryear{Guy et~al.,}{Guy et~al.}{2007}]{Guy2007}
Guy J.,  et~al., 2007, \mn@doi [Astronomy & Astrophysics]
  {10.1051/0004-6361:20066930}, 466, 11

\bibitem[\protect\citeauthoryear{Harris et~al.,}{Harris
  et~al.}{2020}]{Harris2020}
Harris C.~R.,  et~al., 2020, \mn@doi [Nature] {10.1038/s41586-020-2649-2}, 585,
  357

\bibitem[\protect\citeauthoryear{H{\"{o}}flich, Wheeler  \&
  Thielemann}{H{\"{o}}flich et~al.}{1998}]{Hoflich1998}
H{\"{o}}flich P.,  Wheeler J.,   Thielemann F.,  1998, \mn@doi [The
  Astrophysical Journal] {10.1086/305327}, 495, 617

\bibitem[\protect\citeauthoryear{Hunter}{Hunter}{2007}]{Hunter2007}
Hunter J.~D.,  2007, \mn@doi [Computing in Science and Engineering]
  {10.1109/MCSE.2007.55}, 9, 99

\bibitem[\protect\citeauthoryear{Inoue}{Inoue}{2011}]{Inoue2011}
Inoue A.~K.,  2011, \mn@doi [Monthly Notices of the Royal Astronomical Society]
  {10.1111/j.1365-2966.2011.18906.x}, 415, 2920

\bibitem[\protect\citeauthoryear{Jha, Riess  \& Kirshner}{Jha
  et~al.}{2007}]{Jha2007}
Jha S.,  Riess A.~G.,   Kirshner R.~P.,  2007, \mn@doi [The Astrophysical
  Journal] {10.1086/512054}, 659, 122

\bibitem[\protect\citeauthoryear{Kasen, R{\"{o}}pke  \& Woosley}{Kasen
  et~al.}{2009}]{Kasen2009}
Kasen D.,  R{\"{o}}pke F.~K.,   Woosley S.~E.,  2009, \mn@doi [Nature]
  {10.1038/nature08256}, 460, 869

\bibitem[\protect\citeauthoryear{Kelly, Hicken, Burke, Mandel  \&
  Kirshner}{Kelly et~al.}{2010}]{Kelly2010}
Kelly P.~L.,  Hicken M.,  Burke D.~L.,  Mandel K.~S.,   Kirshner R.~P.,  2010,
  \mn@doi [Astrophysical Journal] {10.1088/0004-637X/715/2/743}, 715, 743

\bibitem[\protect\citeauthoryear{Kelsey et~al.,}{Kelsey
  et~al.}{2021}]{Kelsey2021}
Kelsey L.,  et~al., 2021, \mn@doi [Monthly Notices of the Royal Astronomical
  Society] {10.1093/mnras/staa3924}, 501, 4861

\bibitem[\protect\citeauthoryear{Kelsey, Sullivan, Wiseman, Smith, Vincenzi,
  Scolnic  \& Brout}{Kelsey et~al.}{2022}]{Kelsey2022}
Kelsey L.,  Sullivan M.,  Wiseman P.,  Smith M.,  Vincenzi M.,  Scolnic D.,
  Brout D.,  2022, MNRAS, in prep

\bibitem[\protect\citeauthoryear{Kessler \& Scolnic}{Kessler \&
  Scolnic}{2017}]{Kessler2017}
Kessler R.,  Scolnic D.,  2017, \mn@doi [The Astrophysical Journal]
  {10.3847/1538-4357/836/1/56}, 836, 56

\bibitem[\protect\citeauthoryear{Kessler et~al.,}{Kessler
  et~al.}{2009}]{Kessler2009}
Kessler R.,  et~al., 2009, \mn@doi [Astrophysical Journal Supplement Series]
  {10.1088/0067-0049/185/1/32}, 185, 32

\bibitem[\protect\citeauthoryear{Kessler et~al.,}{Kessler
  et~al.}{2019}]{Kessler2019}
Kessler R.,  et~al., 2019, \mn@doi [Monthly Notices of the Royal Astronomical
  Society] {10.1093/mnras/stz463}, 485, 1171

\bibitem[\protect\citeauthoryear{Kewley \& Ellison}{Kewley \&
  Ellison}{2008}]{Kewley2008}
Kewley L.~J.,  Ellison S.~L.,  2008, \mn@doi [The Astrophysical Journal]
  {10.1086/587500}, 681, 1183

\bibitem[\protect\citeauthoryear{Lampeitl et~al.,}{Lampeitl
  et~al.}{2010}]{Lampeitl2010}
Lampeitl H.,  et~al., 2010, \mn@doi [Monthly Notices of the Royal Astronomical
  Society] {10.1111/j.1365-2966.2009.15851.x}, 401, 2331

\bibitem[\protect\citeauthoryear{Mandel, Narayan  \& Kirshner}{Mandel
  et~al.}{2011}]{Mandel2011}
Mandel K.~S.,  Narayan G.,   Kirshner R.~P.,  2011, \mn@doi [The Astrophysical
  Journal] {10.1088/0004-637X/731/2/120}, 731, 120

\bibitem[\protect\citeauthoryear{Mandel, Scolnic, Shariff, Foley  \&
  Kirshner}{Mandel et~al.}{2017}]{Mandel2017}
Mandel K.~S.,  Scolnic D.~M.,  Shariff H.,  Foley R.~J.,   Kirshner R.~P.,
  2017, \mn@doi [The Astrophysical Journal] {10.3847/1538-4357/aa6038}, 842, 93

\bibitem[\protect\citeauthoryear{Marriner et~al.,}{Marriner
  et~al.}{2011}]{Marriner2011}
Marriner J.,  et~al., 2011, \mn@doi [ApJ] {10.1088/0004-637X/740/2/72}, 740, 72

\bibitem[\protect\citeauthoryear{McKinney}{McKinney}{2010}]{Mckinney2010}
McKinney W.,  2010, in PROC. OF THE 9th PYTHON IN SCIENCE CONF. pp 56--61,
  \mn@doi{10.25080/Majora-92bf1922-00a}

\bibitem[\protect\citeauthoryear{M{\"{o}}ller et~al.,}{M{\"{o}}ller
  et~al.}{2022}]{Moller2022}
M{\"{o}}ller A.,  et~al., 2022, eprint arxiv:2201.11142

\bibitem[\protect\citeauthoryear{Moreno-Raya, Moll{\'{a}},
  L{\'{o}}pez-S{\'{a}}nchez, Galbany, V{\'{i}}lchez, Rosell  \&
  Dom{\'{i}}nguez}{Moreno-Raya et~al.}{2016}]{Moreno-Raya2016}
Moreno-Raya M.~E.,  Moll{\'{a}} M.,  L{\'{o}}pez-S{\'{a}}nchez {\'{A}}.~R.,
  Galbany L.,  V{\'{i}}lchez J.~M.,  Rosell A.~C.,   Dom{\'{i}}nguez I.,  2016,
  \mn@doi [The Astrophysical Journal] {10.3847/2041-8205/818/1/l19}, 818, L19

\bibitem[\protect\citeauthoryear{Nicolas et~al.,}{Nicolas
  et~al.}{2021}]{Nicolas2020}
Nicolas N.,  et~al., 2021, \mn@doi [Astronomy & Astrophysics]
  {10.1051/0004-6361/202038447}, 649, A74

\bibitem[\protect\citeauthoryear{Perlmutter et~al.,}{Perlmutter
  et~al.}{1999}]{Perlmutter1999}
Perlmutter S.,  et~al., 1999, \mn@doi [The Astrophysical Journal]
  {10.1086/307221}, 517, 565

\bibitem[\protect\citeauthoryear{Phillips}{Phillips}{1993}]{Phillips1993}
Phillips M.~M.,  1993, \mn@doi [The Astrophysical Journal] {10.1086/186970},
  413, L105

\bibitem[\protect\citeauthoryear{Popovic, Brout, Kessler  \& Scolnic}{Popovic
  et~al.}{2021a}]{Popovic2021a}
Popovic B.,  Brout D.,  Kessler R.,   Scolnic D.,  2021a, eprint
  arxiv:2112.04456

\bibitem[\protect\citeauthoryear{Popovic, Brout, Kessler, Scolnic  \&
  Lu}{Popovic et~al.}{2021b}]{Popovic2021}
Popovic B.,  Brout D.,  Kessler R.,  Scolnic D.,   Lu L.,  2021b, \mn@doi [The
  Astrophysical Journal] {10.3847/1538-4357/abf14f}, 913, 49

\bibitem[\protect\citeauthoryear{Pskovskii}{Pskovskii}{1977}]{Pskovskii1977}
Pskovskii I.~P.,  1977, SvA, 21, 675

\bibitem[\protect\citeauthoryear{Reddy et~al.,}{Reddy et~al.}{2018}]{Reddy2018}
Reddy N.~A.,  et~al., 2018, \mn@doi [The Astrophysical Journal]
  {10.3847/1538-4357/aaa3e7}, 853, 56

\bibitem[\protect\citeauthoryear{Riess, Press  \& Kirshner}{Riess
  et~al.}{1996}]{Riess1996}
Riess A.~G.,  Press W.~H.,   Kirshner R.~P.,  1996, \mn@doi [The Astrophysical
  Journal] {10.1086/178129}, 473, 88

\bibitem[\protect\citeauthoryear{Riess et~al.,}{Riess et~al.}{1998}]{Riess1998}
Riess A.~G.,  et~al., 1998, \mn@doi [The Astronomical Journal]
  {10.1086/300499}, 116, 1009

\bibitem[\protect\citeauthoryear{Rigault et~al.,}{Rigault
  et~al.}{2020}]{Rigault2018}
Rigault M.,  et~al., 2020, \mn@doi [Astronomy & Astrophysics]
  {10.1051/0004-6361/201730404}, 644, A176

\bibitem[\protect\citeauthoryear{Roman et~al.,}{Roman et~al.}{2018}]{Roman2018}
Roman M.,  et~al., 2018, \mn@doi [Astronomy & Astrophysics]
  {10.1051/0004-6361/201731425}, 615, A68

\bibitem[\protect\citeauthoryear{Rose, Rubin, Strolger  \& Garnavich}{Rose
  et~al.}{2021}]{Rose2021}
Rose B.~M.,  Rubin D.,  Strolger L.,   Garnavich P.~M.,  2021, \mn@doi [The
  Astrophysical Journal] {10.3847/1538-4357/abd550}, 909, 28

\bibitem[\protect\citeauthoryear{Salim, Boquien  \& Lee}{Salim
  et~al.}{2018}]{Salim2018}
Salim S.,  Boquien M.,   Lee J.~C.,  2018, \mn@doi [The Astrophysical Journal]
  {10.3847/1538-4357/aabf3c}, 859, 11

\bibitem[\protect\citeauthoryear{Scolnic \& Kessler}{Scolnic \&
  Kessler}{2016}]{Scolnic2016}
Scolnic D.,  Kessler R.,  2016, \mn@doi [The Astrophysical Journal]
  {10.3847/2041-8205/822/2/L35}, 822, L35

\bibitem[\protect\citeauthoryear{Scolnic et~al.,}{Scolnic
  et~al.}{2018}]{Scolnic2018}
Scolnic D.~M.,  et~al., 2018, \mn@doi [The Astrophysical Journal]
  {10.3847/1538-4357/aab9bb}, 859, 101

\bibitem[\protect\citeauthoryear{Scolnic et~al.,}{Scolnic
  et~al.}{2021}]{Scolnic2021}
Scolnic D.,  et~al., 2021, eprint arxiv:2112.03863

\bibitem[\protect\citeauthoryear{Smith et~al.,}{Smith et~al.}{2020}]{Smith2020}
Smith M.,  et~al., 2020, \mn@doi [Monthly Notices of the Royal Astronomical
  Society] {10.1093/mnras/staa946}, 494, 4426

\bibitem[\protect\citeauthoryear{Sullivan et~al.,}{Sullivan
  et~al.}{2010}]{Sullivan2010}
Sullivan M.,  et~al., 2010, \mn@doi [Monthly Notices of the Royal Astronomical
  Society] {10.1111/j.1365-2966.2010.16731.x}, 406, no

\bibitem[\protect\citeauthoryear{Sullivan et~al.,}{Sullivan
  et~al.}{2011}]{Sullivan2011}
Sullivan M.,  et~al., 2011, \mn@doi [Astrophysical Journal]
  {10.1088/0004-637X/737/2/102}, 737, 102

\bibitem[\protect\citeauthoryear{Tomczak et~al.,}{Tomczak
  et~al.}{2014}]{Tomczak2014}
Tomczak A.~R.,  et~al., 2014, \mn@doi [The Astrophysical Journal]
  {10.1088/0004-637X/783/2/85}, 783, 85

\bibitem[\protect\citeauthoryear{Tremonti et~al.,}{Tremonti
  et~al.}{2004}]{Tremonti2004}
Tremonti C.~A.,  et~al., 2004, The Astronomical Journal, 613, 898

\bibitem[\protect\citeauthoryear{Tripp}{Tripp}{1998}]{Tripp1998}
Tripp R.,  1998, Astronomy and Astrophysics, 331, 815

\bibitem[\protect\citeauthoryear{Vincenzi et~al.,}{Vincenzi
  et~al.}{2021}]{Vincenzi2020}
Vincenzi M.,  et~al., 2021, \mn@doi [Monthly Notices of the Royal Astronomical
  Society] {10.1093/mnras/stab1353}, 505, 2819

\bibitem[\protect\citeauthoryear{Virtanen et~al.,}{Virtanen
  et~al.}{2020}]{Virtanen2020}
Virtanen P.,  et~al., 2020, \mn@doi [Nature Methods]
  {10.1038/s41592-019-0686-2}, 17, 261

\bibitem[\protect\citeauthoryear{Wiseman et~al.,}{Wiseman
  et~al.}{2021}]{Wiseman2021}
Wiseman P.,  et~al., 2021, \mn@doi [Monthly Notices of the Royal Astronomical
  Society] {10.1093/mnras/stab1943}, 506, 3330

\bibitem[\protect\citeauthoryear{Yates, Kauffmann  \& Guo}{Yates
  et~al.}{2012}]{Yates2012}
Yates R.~M.,  Kauffmann G.,   Guo Q.,  2012, \mn@doi [Monthly Notices of the
  Royal Astronomical Society] {10.1111/J.1365-2966.2012.20595.X}, 422, 215

\bibitem[\protect\citeauthoryear{Zahid, Yates, Kewley  \& Kudritzki}{Zahid
  et~al.}{2013}]{Zahid2013}
Zahid H.~J.,  Yates R.~M.,  Kewley L.~J.,   Kudritzki R.~P.,  2013, \mn@doi
  [Astrophysical Journal] {10.1088/0004-637X/763/2/92}, 763, 92

\bibitem[\protect\citeauthoryear{Zahid, Dima, Kudritzki, Kewley, Geller, Hwang,
  Silverman  \& Kashino}{Zahid et~al.}{2014}]{Zahid2014}
Zahid H.~J.,  Dima G.~I.,  Kudritzki R.-P.,  Kewley L.~J.,  Geller M.~J.,
  Hwang H.~S.,  Silverman J.~D.,   Kashino D.,  2014, \mn@doi [The
  Astrophysical Journal] {10.1088/0004-637X/791/2/130}, 791, 130

\makeatother
\end{thebibliography}




\appendix

\section{Simulating uncertainties}
\label{sec:app_errors}
\begin{figure}

	\includegraphics[width=.48\textwidth]{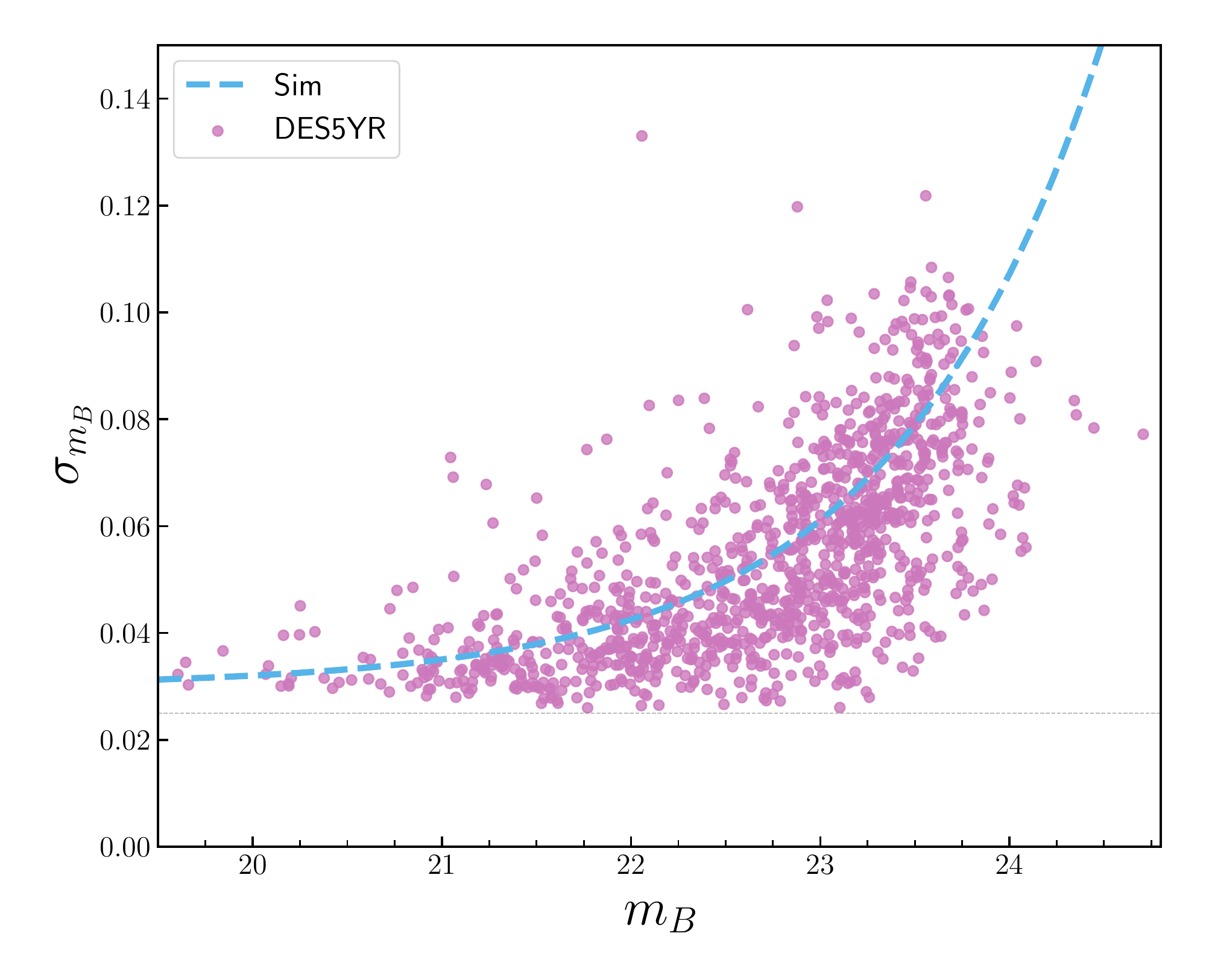}
    \caption{The relationship between peak $B$-band brightness $m_B$ and its uncertainty, $\sigma_{m_B}$. Points are from the DES5YR data and the relationship used in Eq. \ref{eq:mb_err}.}
    \label{fig:mb_errs}
\end{figure}

We simulate uncertainties $\sigma_{m_B}$ on $m_B$ by approximating the relationship between $\sigma_{m_{B},\mathrm{obs}}$ and $m_{B,\mathrm{obs}}$ in the observed data set of K22.

\begin{equation}
     \sigma_{m_B} \sim
     \mathrm{Max}
     \left\{
     \begin{array}{@{}c@{}}
     \mathcal{N}(\overline{\sigma_{m_B}},\sigma_{\sigma_{m_B}}) \\
     0.025
     \end{array}\right. \,,
\end{equation}
where
\begin{equation}
    \overline{\sigma_{m_B}} = 10^{(0.395(m_B-1.5) - 10)}+0.03\,,
    \label{eq:mb_err}
\end{equation}
where $\overline{\sigma_{m_B}}$ is represented by the dashed line in Fig. \ref{fig:mb_errs}. The scatter on the uncertainties $\sigma_{\sigma_{m_B}}$ also increases as a function of $m_B$:
\begin{equation}
    \sigma_{\sigma_{m_B}} = 
    \mathrm{Max}
     \left\{
     \begin{array}{@{}c@{}}
     0.003(m_B-20)\\
     0.003
     \end{array}\right. \,.
\end{equation}
The uncertainty $\sigma_{m_B}$ is calculated after the intrinsic $m_B$ has been adjusted for $x_1$ and $c$ via Eq. \ref{eq:tripp_bs20}. 

\begin{figure}

	\includegraphics[width=.48\textwidth]{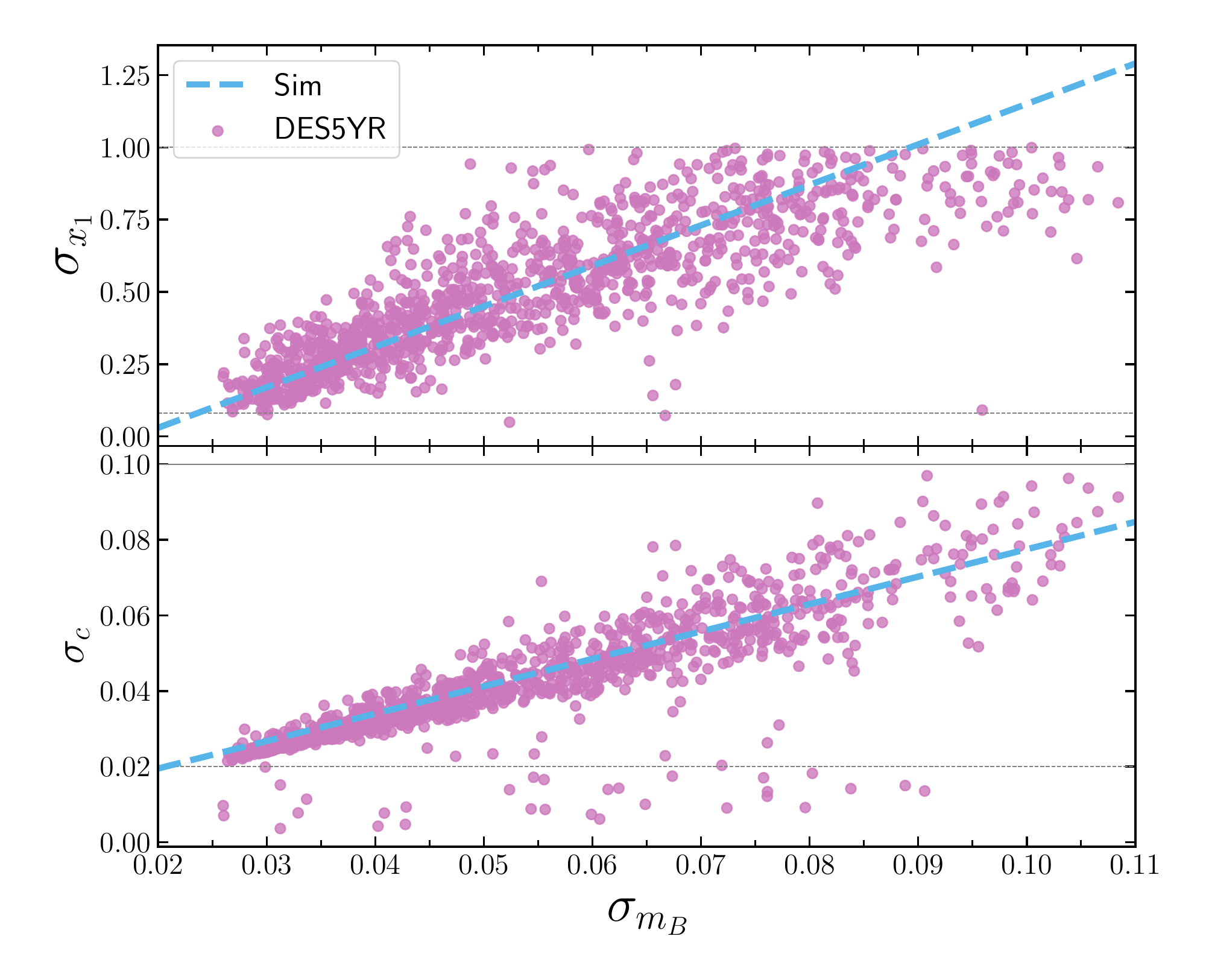}
    \caption{Relationships between $\sigma_{m_B}$ and $\sigma_{x_1}$ (\textit{upper}), and $\sigma_{c}$ \textit{lower}.
    Points are from the DES5YR data and the relationship used in Eq. \ref{eq:mb_err}.}
    \label{fig:x1_c_errs}
\end{figure}

Uncertainties on $x_1$ and $c$ are estimated by linear least squares fits to the DES5YR $\sigma_{x_1} - \sigma_{m_B}$ and $\sigma_{c} - \sigma_{m_B}$ relations respectively. The observed and fitted relationships between the uncertainties are shown in Figs. \ref{fig:mb_errs} \& \ref{fig:x1_c_errs}. We use these uncertainties to add noise to $m_B$, $x_1$, and $c$ by drawing them randomly from Gaussian distributions centred at 0 as per Eq. \ref{eq:sigma_int}.

\section{BBC-BS20 bias corrections}
\label{sec:app_4D}
\begin{figure*}

	\includegraphics[width=\textwidth]{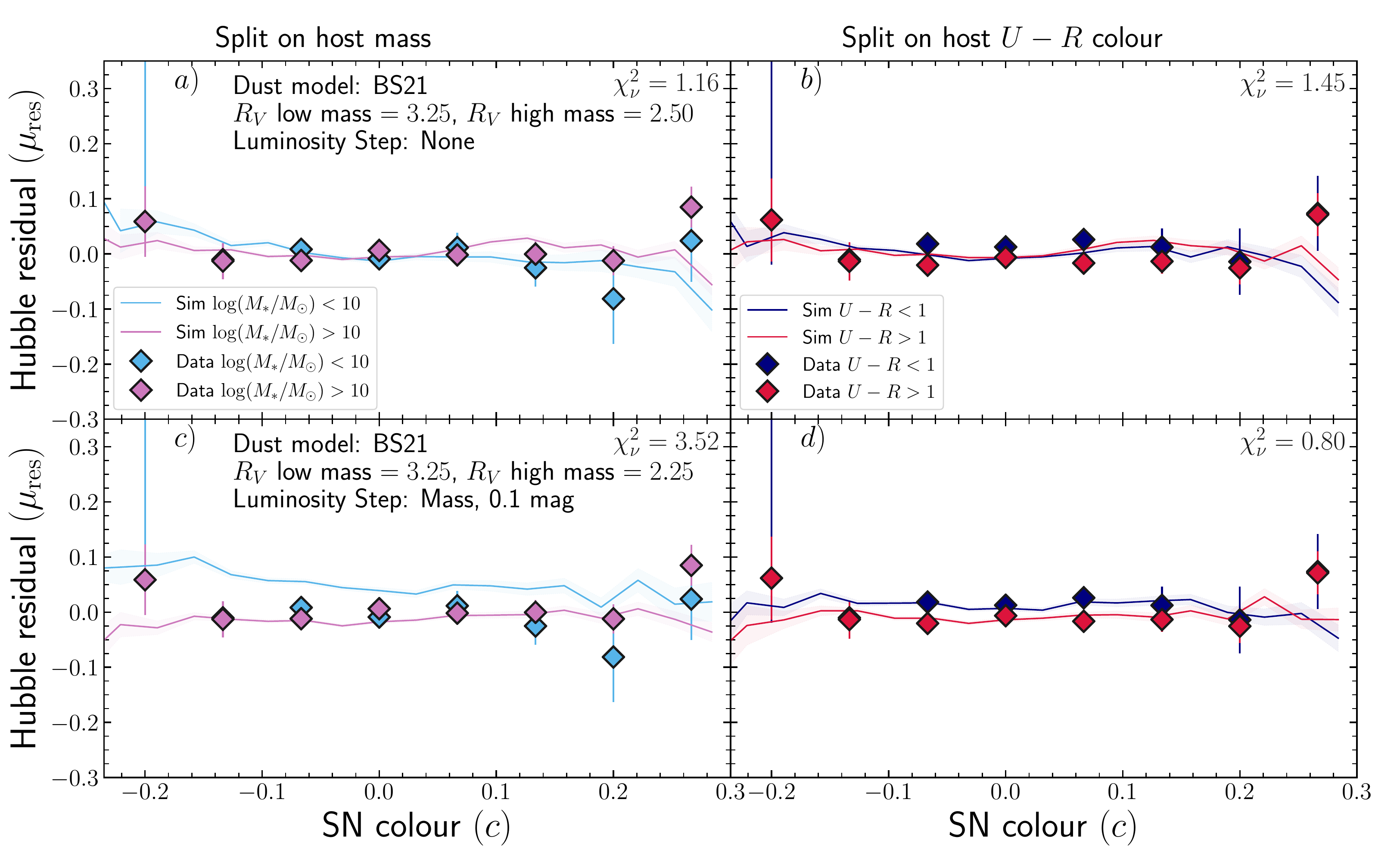}
    \caption{BS21 models as Fig. \ref{fig:HR_c_1D_BS21} but where Hubble residuals for data and simulations have been measured with BBC-BS20. \textit{a):} Simulations from the BS21 model with $\overline{R_V}=3.25$ in low mass host galaxies and $\overline{R_V}=2.5$ in high mass host galaxies; \textit{b):} as \textit{a}, but for the data and simulations split host galaxy $U-R = 1$.
    \textit{c) \& d):} as \textit{a} \& \textit{b} but for the model parameters that best describe the $U-R$ data (right hand panel): $\overline{R_V}=3.25$ in low mass host galaxies and $\overline{R_V}=2.25$ in low mass host galaxies, and an intrinsic luminosity step at $\log(M/\mathrm{M}_{\odot})=10$ of size $0.1$~mag.}
    \label{fig:HR_c_mass_step_4D_BS21}
\end{figure*}

\begin{figure*}

	\includegraphics[width=0.8\textwidth]{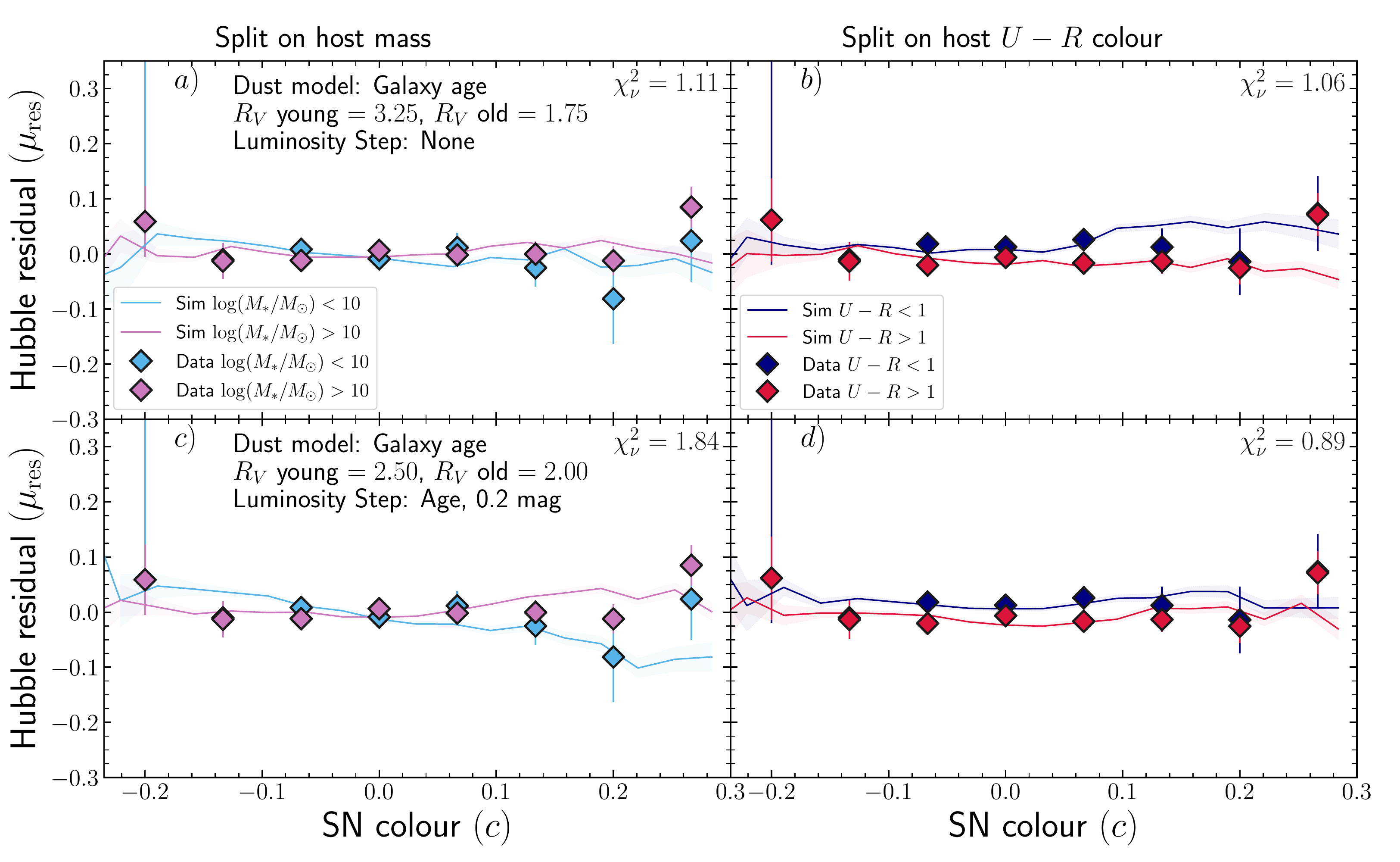}
    \caption{Age-$R_V$ split models as Fig. \ref{fig:HR_c_1D_age}, but where Hubble residuals for data and simulations have been measured with BBC-BS20.}
    \label{fig:HR_c_age_step_4D}
\end{figure*}

\begin{figure*}

	\includegraphics[width=0.8\textwidth]{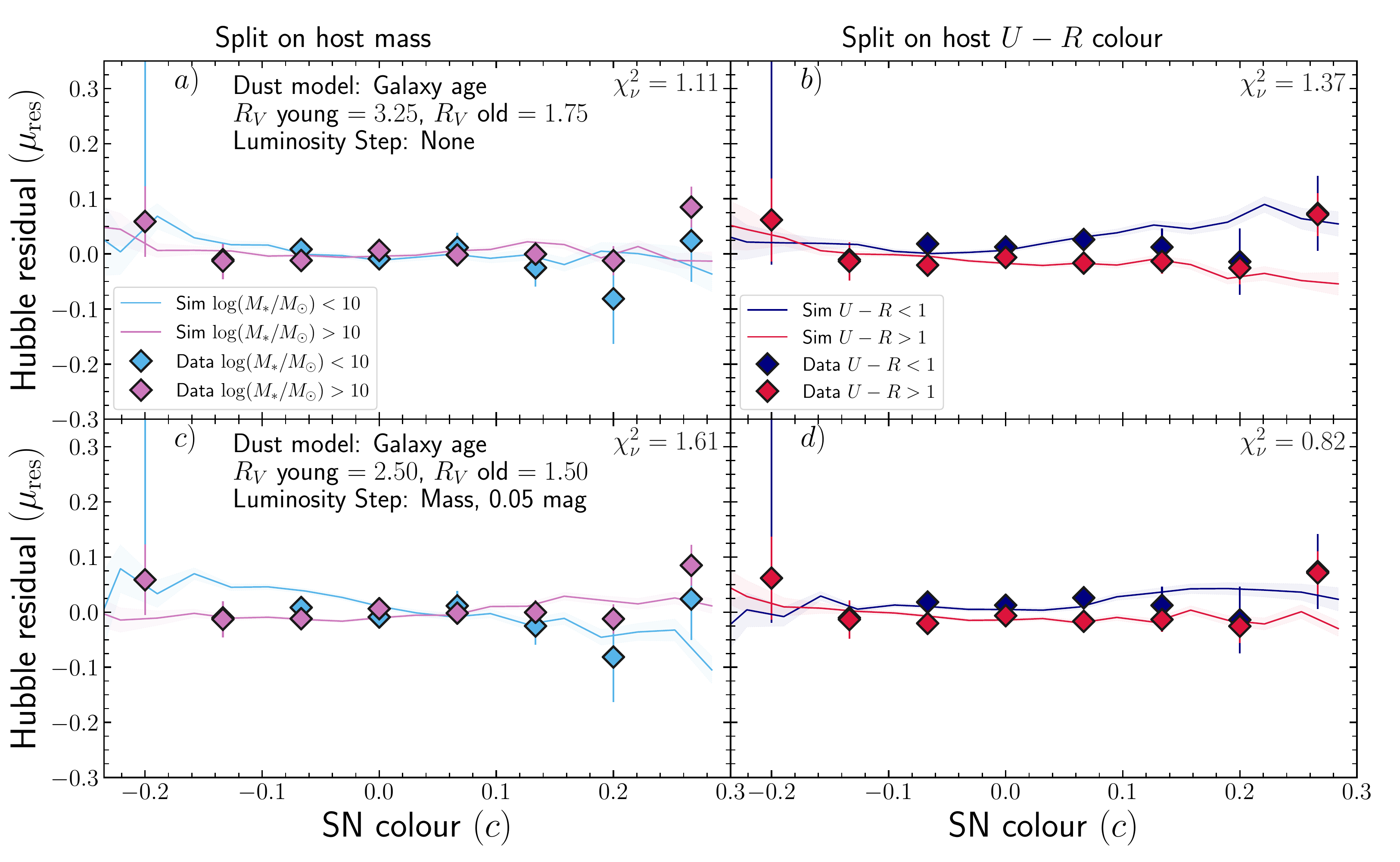}
    \caption{As Fig. \ref{fig:HR_c_mass_step_4D} but for the models best fitting when introducing an intrinsic stellar mass step instead of a SN age step.}
    \label{fig:HR_c_mass_step_4D}
\end{figure*}

Here we present results in the same was as Section \ref{sec:results} but with Hubble residuals measured using a BBC-BS20 bias correction, which inherently assumes $R_V$ changing with stellar mass. The models shown in Figs. \ref{fig:HR_c_mass_step_4D_BS21}-\ref{fig:HR_c_mass_step_4D} correspond to the best match parameters when splitting on stellar mass (blue) and $U-R$ (green), which are presented in Tables \ref{tab:res_split_mass_4D} and \ref{tab:res_split_UR_4D} respectively.
\newpage

\begin{table*}
	\centering
	\caption{Dust population parameters for the  models presented in this work when compared to DES5YR data split at host $\log(M_*/\mathrm{M}_{\odot})=10$ and fit with BBC-BS20 (left hand panels of Figs. \ref{fig:HR_c_mass_step_4D_BS21}-\ref{fig:HR_c_mass_step_4D}.)}
	\label{tab:res_split_mass_4D}
	\begin{tabular}{lcccccccccccccc} 
	    \hline
		Model Name &BiasCor& $\overline{R_{V,1}}$  & $\sigma_{R_V, 1}$  &  $\overline{R_{V,2}}$  & $\sigma_{R_V, 2}$  & $Y$ & $Y_{\mathrm{split}}$ & $Y_1$  & $Y_2$ &$X$ & $X_{\mathrm{split}}$& $\gamma_{X}$ & $\chi^2_{\nu,\mu_{\mathrm{res}}}(U-R)$ \\
		\hline
		BS21 & BS20 & 2.5 & 1.0 & 3.25 & 1.0 &   $M_*$  &  $10^{10}~ \mathrm{M}_{\odot}$ & - & -& $M_*$&$10^{10}~ \mathrm{M}_{\odot}$ & 0.00 & 1.16\\
		BS21 & BS20 & 2.25 & 1.0 & 3.0 & 1.0 &   $M_*$  &  $10^{10}~ \mathrm{M}_{\odot}$ & - & -&$\tau_{\mathrm{A}}$&0.75~Gyr  & 0.05 & 1.23\\
		Age $R_V$& BS20 & 1.75 & 1.0 & 3.25 & 1.0 & $\tau_{\mathrm{G}}$ &  3~Gyr& - & - &$\tau_{\mathrm{A}}$&0.75~Gyr  & 0.0 &1.11\\
		Age $R_V$& BS20 & 1.75 & 1.0 & 3.25 & 1.0 & $\tau_{\mathrm{G}}$ &  3~Gyr& - & - &$M_*$&$10^{10}~ \mathrm{M}_{\odot}$ & 0.0 &1.11\\
		Age $R_V$ Linear & BS20 &1.5 & 1.0 & 3.75 & 1.0 & $\tau_{\mathrm{A}}$ & - & 0.1~Gyr & 10~Gyr &$\tau_{\mathrm{A}}$ & 0.75~Gyr& 0.20 & 1.40\\
		\hline
	\end{tabular}
\end{table*}

\begin{table*}
	\centering
	\caption{Dust population parameters for the  models presented in this work when compared to DES5YR data split at host $U-R=1$ and fit with BBC-BS20 (right hand panels of Figs. \ref{fig:HR_c_mass_step_4D_BS21}-\ref{fig:HR_c_mass_step_4D}.)}
	\label{tab:res_split_UR_4D}
	\begin{tabular}{lcccccccccccccc} 
	    \hline
		Model Name &BiasCor& $\overline{R_{V,1}}$  & $\sigma_{R_V, 1}$  &  $\overline{R_{V,2}}$  & $\sigma_{R_V, 2}$  & $Y$ & $Y_{\mathrm{split}}$ & $Y_1$  & $Y_2$ &$X$ & $X_{\mathrm{split}}$& $\gamma_{X}$ & $\chi^2_{\nu,\mu_{\mathrm{res}}}(U-R)$ \\
		\hline
		BS21 & BS20 & 2.25 & 1.0 & 3.25 & 1.0 &   $M_*$  &  $10^{10}~ \mathrm{M}_{\odot}$ & - & -& $M_*$&$10^{10}~ \mathrm{M}_{\odot}$ & 0.1 & 0.80\\
		BS21 & BS20 & 1.75 & 1.0 & 3.75 & 1.0 &   $M_*$  &  $10^{10}~ \mathrm{M}_{\odot}$ & - & -&$\tau_{\mathrm{A}}$&0.75~Gyr  & 0.05 & 0.76\\
		Age $R_V$& BS20 & 2.0 & 1.0 & 2.5 & 1.0 & $\tau_{\mathrm{G}}$ &  3~Gyr& - & - &$\tau_{\mathrm{A}}$&0.75~Gyr  & 0.15 &0.89\\
		Age $R_V$& BS20 & 1.5 & 1.0 & 2.5 & 1.0 & $\tau_{\mathrm{G}}$ &  3~Gyr& - & - &$M_*$&$10^{10}~ \mathrm{M}_{\odot}$& 0.05 &0.82\\
		Age $R_V$ Linear & BS20 &1.5 & 1.0 & 3.75 & 1.0 & $\tau_{\mathrm{A}}$ & - & 0.1~Gyr & 10~Gyr &$\tau_{\mathrm{A}}$ & 0.75~Gyr& 0.20 & 1.22\\	
		
		\hline
	\end{tabular}
\end{table*}

\onecolumn
\parbox{\textwidth}{
$^{1}$ School of Physics and Astronomy, University of Southampton,  Southampton, SO17 1BJ, UK\\
$^{2}$ Department of Physics, Duke University Durham, NC 27708, USA\\
$^{3}$ Institute of Cosmology and Gravitation, University of Portsmouth, Portsmouth, PO1 3FX, UK\\
$^{4}$ Center for Astrophysics $\vert$ Harvard \& Smithsonian, 60 Garden Street, Cambridge, MA 02138, USA\\
$^{5}$ Einstein Fellow\\
$^{6}$ School of Mathematics and Physics, University of Queensland,  Brisbane, QLD 4072, Australia\\
$^{7}$ Institut d'Estudis Espacials de Catalunya (IEEC), 08034 Barcelona, Spain\\
$^{8}$ Institute of Space Sciences (ICE, CSIC),  Campus UAB, Carrer de Can Magrans, s/n,  08193 Barcelona, Spain\\
$^{9}$ Centre for Gravitational Astrophysics, College of Science, The Australian National University, ACT 2601, Australia\\
$^{10}$ The Research School of Astronomy and Astrophysics, Australian National University, ACT 2601, Australia\\
$^{11}$ Centre for Astrophysics \& Supercomputing, Swinburne University of Technology, Victoria 3122, Australia\\
$^{12}$ Univ Lyon, Univ Claude Bernard Lyon 1, CNRS, IP2I Lyon / IN2P3, IMR 5822, F-69622 Villeurbanne, France\\
$^{13}$ Laborat\'orio Interinstitucional de e-Astronomia - LIneA, Rua Gal. Jos\'e Cristino 77, Rio de Janeiro, RJ - 20921-400, Brazil\\
$^{14}$ Fermi National Accelerator Laboratory, P. O. Box 500, Batavia, IL 60510, USA\\
$^{15}$ Department of Physics, University of Michigan, Ann Arbor, MI 48109, USA\\
$^{16}$ CNRS, UMR 7095, Institut d'Astrophysique de Paris, F-75014, Paris, France\\
$^{17}$ Sorbonne Universit\'es, UPMC Univ Paris 06, UMR 7095, Institut d'Astrophysique de Paris, F-75014, Paris, France\\
$^{18}$ University Observatory, Faculty of Physics, Ludwig-Maximilians-Universit\"at, Scheinerstr. 1, 81679 Munich, Germany\\
$^{19}$ Department of Physics \& Astronomy, University College London, Gower Street, London, WC1E 6BT, UK\\
$^{20}$ Kavli Institute for Particle Astrophysics \& Cosmology, P. O. Box 2450, Stanford University, Stanford, CA 94305, USA\\
$^{21}$ SLAC National Accelerator Laboratory, Menlo Park, CA 94025, USA\\
$^{22}$ Instituto de Astrofisica de Canarias, E-38205 La Laguna, Tenerife, Spain\\
$^{23}$ Universidad de La Laguna, Dpto. Astrofísica, E-38206 La Laguna, Tenerife, Spain\\
$^{24}$ Center for Astrophysical Surveys, National Center for Supercomputing Applications, 1205 West Clark St., Urbana, IL 61801, USA\\
$^{25}$ Department of Astronomy, University of Illinois at Urbana-Champaign, 1002 W. Green Street, Urbana, IL 61801, USA\\
$^{26}$ Institut de F\'{\i}sica d'Altes Energies (IFAE), The Barcelona Institute of Science and Technology, Campus UAB, 08193 Bellaterra (Barcelona) Spain\\
$^{27}$ Astronomy Unit, Department of Physics, University of Trieste, via Tiepolo 11, I-34131 Trieste, Italy\\
$^{28}$ INAF-Osservatorio Astronomico di Trieste, via G. B. Tiepolo 11, I-34143 Trieste, Italy\\
$^{29}$ Institute for Fundamental Physics of the Universe, Via Beirut 2, 34014 Trieste, Italy\\
$^{30}$ Hamburger Sternwarte, Universit\"{a}t Hamburg, Gojenbergsweg 112, 21029 Hamburg, Germany\\
$^{31}$ Department of Physics, IIT Hyderabad, Kandi, Telangana 502285, India\\
$^{32}$ Jet Propulsion Laboratory, California Institute of Technology, 4800 Oak Grove Dr., Pasadena, CA 91109, USA\\
$^{33}$ Institute of Theoretical Astrophysics, University of Oslo. P.O. Box 1029 Blindern, NO-0315 Oslo, Norway\\
$^{34}$ Kavli Institute for Cosmological Physics, University of Chicago, Chicago, IL 60637, USA\\
$^{35}$ Instituto de Fisica Teorica UAM/CSIC, Universidad Autonoma de Madrid, 28049 Madrid, Spain\\
$^{36}$ Department of Physics and Astronomy, University of Pennsylvania, Philadelphia, PA 19104, USA\\
$^{37}$ Observat\'orio Nacional, Rua Gal. Jos\'e Cristino 77, Rio de Janeiro, RJ - 20921-400, Brazil\\
$^{38}$ Santa Cruz Institute for Particle Physics, Santa Cruz, CA 95064, USA\\
$^{39}$ Center for Cosmology and Astro-Particle Physics, The Ohio State University, Columbus, OH 43210, USA\\
$^{40}$ Department of Physics, The Ohio State University, Columbus, OH 43210, USA\\
$^{41}$ Instituci\'o Catalana de Recerca i Estudis Avan\c{c}ats, E-08010 Barcelona, Spain\\
$^{42}$ Physics Department, 2320 Chamberlin Hall, University of Wisconsin-Madison, 1150 University Avenue Madison, WI  53706-1390\\
$^{43}$ Department of Astronomy, University of California, Berkeley,  501 Campbell Hall, Berkeley, CA 94720, USA\\
$^{44}$ Institute of Astronomy, University of Cambridge, Madingley Road, Cambridge CB3 0HA, UK\\
$^{45}$ Department of Astrophysical Sciences, Princeton University, Peyton Hall, Princeton, NJ 08544, USA\\
$^{46}$ Department of Physics and Astronomy, Pevensey Building, University of Sussex, Brighton, BN1 9QH, UK\\
$^{47}$ Centro de Investigaciones Energ\'eticas, Medioambientales y Tecnol\'ogicas (CIEMAT), Madrid, Spain\\
$^{48}$ Computer Science and Mathematics Division, Oak Ridge National Laboratory, Oak Ridge, TN 37831\\
$^{49}$ Excellence Cluster Origins, Boltzmannstr.\ 2, 85748 Garching, Germany\\
$^{50}$ Max Planck Institute for Extraterrestrial Physics, Giessenbachstrasse, 85748 Garching, Germany\\
$^{51}$ Universit\"ats-Sternwarte, Fakult\"at f\"ur Physik, Ludwig-Maximilians Universit\"at M\"unchen, Scheinerstr. 1, 81679 M\"unchen, Germany\\
}

\bsp	
\label{lastpage}
\end{document}